\documentclass[12pt,preprint]{aastex}
\newcommand {\boom}{{\sc Boom\-erang} }
\newcommand {\boomn}{{\sc Boom\-erang}}

\slugcomment{Submitted to Astrophysical Journal 6/14/02, Revised 12/13/02}

\begin{document}

\title{\boom: A Balloon-borne Millimeter Wave Telescope and Total Power
Receiver for Mapping Anisotropy in the Cosmic Microwave Background} 

\author{
B.P. Crill\altaffilmark{1,2}, P.A.R. Ade\altaffilmark{3},
D. R. Artusa\altaffilmark{1}, R.S. Bhatia\altaffilmark{1},
J.J. Bock\altaffilmark{1,4}, A. Boscaleri\altaffilmark{5}, P. 
Cardoni\altaffilmark{6},
S. E. Church\altaffilmark{7}, K. Coble\altaffilmark{8,9},
P. deBernardis\altaffilmark{10},
G. deTroia\altaffilmark{10},
P. Farese\altaffilmark{11}, K. M. Ganga\altaffilmark{12}, M. 
Giacometti\altaffilmark{7},
C. V. Haynes\altaffilmark{13}, E. Hivon\altaffilmark{1,12},
V.V. Hristov\altaffilmark{1}, 
A. Iacoangeli\altaffilmark{6}, W.C. Jones\altaffilmark{1},A.E. 
Lange\altaffilmark{1},
L. Martinis\altaffilmark{7}, S. Masi\altaffilmark{10},
P.V. Mason\altaffilmark{1,4},P.D. Mauskopf\altaffilmark{3},
L. Miglio\altaffilmark{14},  T. Montroy\altaffilmark{11},
C.B. Netterfield\altaffilmark{14}, C. G. Paine\altaffilmark{1,4},
E. Pascale\altaffilmark{5}, F. Piacentini\altaffilmark{10},
G. Polenta\altaffilmark{9},
F. Pongetti\altaffilmark{15}, G. Romeo\altaffilmark{15},
J.E. Ruhl\altaffilmark{16}, F. Scaramuzzi\altaffilmark{7},
D. Sforna\altaffilmark{6}, and A.D. Turner\altaffilmark{4}. 
}

\affil{
 $^{1}$ Observational Cosmology, MS 59-33, California Institute of  
Technology, Pasadena, CA 91125, USA \\
 $^{2}$ Physics Department, California State University - Dominguez Hills, 
Carson, CA 90747, USA (Current Affilliation) \\
 $^{3}$ Department of Physics and Astronomy, Cardiff University, Cardiff 
CF24 3YB, UK\\
 $^{4}$ Jet Propulsion Laboratory, Pasadena, CA, USA \\
 $^{5}$ IROE-CNR, Firenze, Italy \\
 $^{6}$ ENEA, Frascati, Italy \\
 $^{7}$ Department of Physics, Stanford Univ., Palo Alto, CA, USA \\
 $^{8}$ University of Chicago Dept. of Astronomy and Astrophysics, Chicago, IL, USA\\
 $^{9}$ Adler Planetarium and Astronomy Museum, Chicago, IL, USA\\
 $^{10}$ Dipartimento di Fisica, Universit\'{a} La Sapienza, Roma, Italy \\
 $^{11}$ Department of Physics, Univ. of California, Santa Barbara, CA, USA \\
 $^{12}$ Infrared Processing and Analysis Center, Pasadena, CA, USA\\
 $^{13}$ Queen Mary and Westfield College, London, UK \\
 $^{14}$ Department of Astronomy, University of Toronto, 60 George St, 
Toronto ON M5S 3H8, Canada \\
 $^{15}$ Istituto Nazionale di Geofisica, Roma, Italy \\
 $^{16}$ Physics Department, Case Western Reserve University, Cleveland, OH, USA \\
}

\begin{abstract}
We describe \boomn; a balloon-borne microwave telescope
designed to map the Cosmic Microwave Background (CMB) at a resolution of
10$'$ from the Long Duration Balloon (LDB) platform.   The millimeter-wave
receiver employs new technology in bolometers, readout electronics, cold
re-imaging optics, millimeter-wave filters, and cryogenics to obtain high
sensitivity to CMB anisotropy.  Sixteen detectors observe
in 4 spectral bands centered at 90, 150, 240 and 410 GHz. The wide
frequency coverage, the long duration flight, the optical design and
the observing strategy provide strong rejection of
systematic effects.  We report the flight performance of  the instrument
during a 10.5 day stratospheric balloon flight launched from McMurdo
Station, Antarctica that mapped $\sim$ 2000 square  degrees of the sky.  
\end{abstract}


\section{Introduction}

Measurements of the angular power spectrum of anisotropy in the Cosmic
Microwave Background (CMB) are greatly enhancing our 
knowledge of fundamental properties of the universe.   In particular,
models of the early universe predict the existence of a series 
of harmonic peaks in the angular power spectrum at degree scales.  The
precise determination of the amplitude and angular scale of the peaks
provides strong constraints on these models and enables their
parameters to be determined with great precision
(\cite{sachswolfe}; \cite{hu}).  After the discovery 
of the large scale anisotropies by COBE-DMR (\cite{Bennett}), 
a host of ground-based and balloon-borne  observations detected 
a first peak at a multipole moment of $\ell \sim$ 200 in
the angular power spectrum of the CMB (\cite{bjk};\cite{MAT};\cite{b97}). 
Here we describe the \boom experiment, which has made a deep map of the 
CMB at sub-degree resolution and a precision measurement of the angular 
scale and amplitude of the first peak (\cite{pdb}; \cite{pdb02}) from the Long  Duration
Balloon (LDB) platform.    

The \boom instrument consists of a 1.3-m off-axis telescope that feeds
a bolometric array receiver.  The receiver is housed inside a long
duration liquid helium cryostat. A sorption pumped $^3$He refrigerator
maintains the detectors at 280 mK. Observations are made in 4 spectral
bands centered at 90, 150, 240, and 410 GHz (3mm, 2mm, 1.3mm, and 750
$\mu$m) with angular resolutions of 18$'$, 10$'$, 14$'$, and 13$'$
FWHM respectively. A test flight of the \boom payload in a different
configuration (\cite{b97inst}) flew in a 6 hour
engineering flight from Palestine, Texas in 1997 (\cite{b97};
\cite{b972}).  The configuration of the instrument described here flew
in a 252 hour flight from McMurdo Station, Antarctica in 1998-1999.

\section{Instrument Overview}

\boom incorporates four unique design features that allow a
precision measurement of the angular power spectrum of the CMB.    
First, \boom is a quasi total-power radiometer.  The temperature of one
part of the sky is measured relative to its surroundings by slowly (scan 
period 1-2 minutes) scanning the entire telescope in azimuth.  The output
from each detector is AC coupled to an amplifier.  The scan appears at
very  low frequency (0.008 - 0.016 Hz).  The stable, virtually transparent
atmosphere at balloon float altitude and the intrinsic stability of the
bolometric detectors and readout amplifier chain make it possible to map
large areas of the sky with high sensitivity.

Second, \boom is designed to take advantage of the long integration time
possible from a balloon borne platform flown over the Antarctic.  During 
the austral summer, the polar vortex winds provide a stable orbit for
balloons at the top of the stratosphere ($\sim$ 38 km altitude) at a distance of 1200~km from
the pole.  This observation platform provides flight durations of 7 to
20 days, thus allowing  measurements to be repeated many times in order
to check for systematic effects.  The relatively small fraction
($<$10\%) of the sky that is accessible from a balloon platform during
the austral summer fortuitously includes the part of the sky that is
lowest in foreground contamination (\cite{dustmap}). 

Third, the \boom receiver has a high instantaneous sensitivity, due to
its optimized low-background bolometers and high-bandwidth feeds
operating at cryogenic temperatures.  The channels at 90 GHz and 150 GHz
are positioned in frequency to optimally avoid galactic foreground
contamination.  Combining these channels with those at 240 GHz and 410
GHz allows powerful detection and removal of foreground signals.  

Finally, the \boom receiver uses re-imaging optics that provide 
excellent image quality over a large focal plane.   The receiver
simultaneously measures 16 bolometer channels in 8 pixels with beams
separated by up to $4^{\circ}$ on the sky.  The wide format of the focal
plane and large number of detectors allowed by the re-imaging optics
provides the ability to detect and remove temporally correlated
noise, since observations of a specific region of the sky by different
detectors are well-separated in time.

\section{Instrument Description}
\subsection{Detectors}

\boom uses silicon nitride micromesh ``spider web'' bolometers
(\cite{bolo}) that were developed specifically for use in an
environment with a high  cosmic ray flux, such as above the Earth's
polar regions. The bolometer consists of a finely photolithographed mesh
that provides high absorption efficiency over a wide band with low heat
capacity and cosmic  ray cross section (Figure \ref{fig:bolopic}).
A Neutron Transmutation Doped  (NTD)  germanium thermistor provides high
sensitivity and extremely good 
stability.  The fundamental limit of the sensitivity of a bolometer is
phonon noise in the thermal link between the absorber and the
heat sink.  In this case the Noise Equivalent Power, $NEP =
\gamma\sqrt{4k_BT^2G}$ where $G$ is the thermal  conductance, $T$ is the
bath  temperature, and $\gamma \sim 1.2$ takes
into account the contribution from Johnson noise in the NTD Ge
thermistor.  For a given background load $Q$, maximum sensitivity is
achieved for $G \sim Q/T$ (\cite{mather}).    The desired  time constant
of a bolometer $\tau \sim C/G$ (where $C$ is the heat capacity of the 
bolometer) for a given modulation scheme sets a
limit on the minimum thermal conductance that can be selected.   The
bolometers are optimized for a 10~K Rayleigh-Jeans spectrum background
(\cite{boloopt}).  

\subsection{Readout Electronics}

The slow scan observation scheme requires stability of the
detector and the readout electronics, from the bolometer thermal cutoff
frequency ($\sim$ 10 Hz) down to the characteristic scan frequency at
tens of mHz. The detector readout scheme implements an electronic
modulation/demodulation technique to provide stability at low
frequency, by moving the signal bandwidth well above the 1/f knee of the 
JFET and the warm amplifer.  Modulation is achieved by biasing the bolometer
with an AC current.

The detectors are AC voltage biased with a 318~Hz sine wave.  Dual
10M$\Omega$ load resistors provide an approximate current bias. The
signal from each detector passes through a JFET source follower circuit
on the 2~K stage, lowering the output impedance and reducing the
susceptibility to  microphonics. The JFETs were selected for low power
dissipation and  are  packaged by IR Labs (IR Labs TIA). 

The signal from the cold JFETs passes through a preamplifier stage and a
band pass filter stage prior to reaching the synchronous detector.

The synchronous demodulator (lock-in) multiplies the signal
synchronously by +1/-1 times the bias reference. A 4-pole  
Butterworth low pass filter with cutoff at 20~Hz  removes the
high-frequency terms of the product and acts as an anti-aliasing
filter for the Data Acquisition System (DAS).  An AC-coupling high pass
filter with a cutoff at 0.016~Hz and extra gain is applied to the signal
to match the signal dynamics to the DAS dynamic range.  The total gain of 
the amplifier chain is 5$\times$10$^{4}$.

The entire readout electronics contribute less than 10 $nV_{rms}/\sqrt{Hz}$
to the signal noise in the frequency range of 0.016 Hz to 20 Hz.  The
AC-coupled signals are sampled by the DAS at 60~Hz.   Figure \ref{fig:readout} 
shows a block diagram of the analog readout electronics. 

The DAS is constructed entirely from discrete CMOS logic, with the data
frame defined by a UV-EPROM. The 38 kbps bi-phase output is
sent to a line-of-sight  transmitter and to the data  storage system.
The data stream is also stored on board by a pair of redundant commercial
grade 486DX/4 PC's to commercial hard disks and to a Digital Audio Tape (DAT).
The PC's 
additionally compress the data and send it over a 4 kilobit s$^{-1}$ TDRSS 
satellite link to the ground station for in-flight viewing.  The PC's are
housed in a pressure vessel to maintain 1 atmosphere of pressure for the 
hard disks.
 
\subsection{Rejection of RF and Microphonic Interference}

Bolometric detectors are susceptible to spurious heating from radio frequency
interference (RFI) dissipated in the thermistor or to pickup from 
microphonic response of
the bolometer or  wiring.   This is especially of concern at the high
impedance portion of the wiring between the bolometer and the cold JFET
followers.  There are several microwave transmitters for satellite
communcations on board the  \boom payload that are potentially sources 
of RF pickup in the bolometers.  The frequencies of concern are an ARGOS
400 MHz transmitter and a TDRSS 2.3 GHz transmitter.

To keep RFI power from reaching the detectors, each bolometer cavity is 
constructed as a Faraday cage.  The only entrances into this cage are
through the feed horn and via the wiring to the detector.  The feed horn
has a waveguide cutoff at the lower edge of the band, preventing any RFI 
from reaching the detector via that path. The wires that exit 
the cavity are connected to ground through surface mount 20 pF capacitors 
that  
provide protection from RFI close to the detector.   All bolometer leads
from the JFET preamp stage to the warm amplifiers run through cast
eccosorb filters.  The cast eccosorb filters consist of stripline cables 
potted in cast eccosorb (EV Roberts CR-124) and have significant
attenuation above a few~GHz.  These provide the only exit from a second
Faraday cage surrounding the 2~K stage.

A third Faraday cage surrounds the bolometers, room temperature amplifers,
cryostat electronics and wiring.  The amplified signals exit the third
Faraday cage through Spectrum $\Pi$ filters (Spectrum 1212-0502) to the DAS.

The low suspended mass of the micromesh bolometers naturally provides a
reduction in microphonic response (measured resonant frequency $>$1
kHz). The high impedance wiring connecting the bolometers to the JFET follower stage is carefully tied down. The wiring
on the 0.3~K stage consists of 27 AWG shielded twisted pair that is 
strapped down with teflon tape and nylon cord.  Between the 0.3~K and
2~K stages, cables made with low thermal conductivity 50~$\mu$m manganin
wire are strapped to the Vespel posts that support the cold stage.
Inside the JFET box, all high impedance wires are potted in silicone RTV. 

\subsection{Cryostat}

The cryogenic system for \boom keeps the detectors at their operating
temperature  of 0.28~K for the entire two weeks of an LDB balloon flight.
It has 60 liters of volume at 2~K to contain the photometers, re-imaging 
optics, baffles, and cold preamplifiers. The system is composed of a
self contained sorption-pumped $^3$He  refrigerator (\cite{cryo1}) and
a helium/nitrogen main  cryostat (\cite{cryo2}). The $^3$He fridge 
contains 34 
liters STP of $^3$He and runs at 0.280~K with a load of 27 $\mu$W.  The
main cryostat holds 65 liters of liquid nitrogen and 69 liters of liquid
helium. The tanks, toroidal in shape, are suspended by Kevlar
cords. Radiation from 300~K to the 77~K tank is shielded by a blanket 
of 30 layers of aluminized mylar superinsulation.  Radiation from the 
77~K tank to the helium tank is shielded by an intermediate temperature
shield cooled to $<$ 20~K by the boiloff from the helium tank. External
plumbing with electrically actuated valves is used to  pressurize the
nitrogen bath during the flight and to maintain the low pressure of the
helium bath during ascent.  The helium bath is vented to ambient
pressure at float altitude to maintain the lowest possible temperature.

\subsection{Optics}

The \boom telescope (Figure \ref{fig:optics}) consists of an ambient 
temperature (-20$^{\circ}$ C in
the shade at float altitude) off-axis parabolic primary mirror that 
feeds a pair of cold re-imaging mirrors inside the cryostat.  The
primary mirror is 1.3~m in diameter and has a $45^{\circ}$ off-axis 
angle for a projected size of 1.3$\times$1.2~m. The mirror and receiver 
can be tipped in
elevation by $+10^{\circ}$ and  
$-12^{\circ}$ to cover elevation angles from $33^{\circ}$ to
$55^{\circ}$.  Radiation from the sky is reflected by the primary mirror
and passes into the cryostat through a thin (50 $\mu m$) polypropylene
vacuum window near the prime focus.  The window is divided in two circles side
by side, each 6.6 cm in diameter.  Filters to reject high frequency
radiation and to reduce the thermal load on the 2~K and 0.3~K stages of
the cryostat are mounted on the 77~K and 2~K shields in front of the
cold mirrors.  These are capacitive multilayer metal mesh low pass
filters with cutoffs at 540 GHz and 480 GHz respectively.  Neutral
density filters with transmission of 1.5\% used to reduce the loading on
the detectors for ground testing are mounted on a mechanism that can
move the filters in and out of the beam near the prime focus.  

Off-axis ellipsoidal secondary and tertiary mirrors
surrounded by black baffles re-image the prime focus onto a detector
focal plane with diffraction-limited performance at 1~mm over a 
$2^{\circ} \times 5^{\circ}$ field of view.   

The re-imaging optics form an image of the primary mirror at the 10~cm
diameter tertiary mirror.  The size of the tertiary mirror 
limits the geometric illumination of the primary mirror to the central
50\% in area to reduce sidelobe response.  

The secondary and tertiary mirrors are ellipsoidal reflectors with
effective focal lengths of 20~cm and 33~cm respectively.  The mirrors have
been optimized with CodeV software (\cite{pdmthesis}). The prime focus
is fed from the illuminated portion of the primary at f/2, so the detector 
focal plane is fed at f/3.3.  The
tertiary mirror is 10~cm in diameter, corresponding to an 85~cm diameter
aperture on the 1.3~m diameter primary. 

The \boom focal plane contains four single-frequency
channels fed by smooth-walled conical horns and four multicolor
photometers fed by parabolic concentrators (Winston horns).  Although
the image quality from the optics is diffraction limited at 150 GHz over
a $2^{\circ} \times 5^{\circ}$ field, all of the feed optics are placed
inside two circles $2^{\circ}$ in diameter, separated center to center
by  $3.5^{\circ}$.  The focal plane area outside these circles is
unusable because it is vignetted by blocking filters at the entrance to
the 2~K optics box and on the 77~K shield.  The horn positions are
optimized using geometric ray tracing.  All of the feeds are oriented
towards the center of the tertiary mirror.  The phase center of the
conical horns and the center of the aperture of the parabolic horns are
placed at the focus.

A schematic of the relative positions and sizes of the beams on the sky
is shown in Figure \ref{fig:focalmap}.  The focal plane layout is
chosen so as to repeat observations of the same part of the sky on many
different time scales.  The telescope scans in azimuth, and at
1 degree per second (dps) at $45^{\circ}$ elevation, channels on opposite sides of
the focal plane will observe the same sky $\sim$ 2~s apart.  The
rotation of the sky above Antarctica allows each row of detectors to
observe the same patch of sky $\sim$ 30~minutes apart.  

\subsection{Focal Plane}
\subsubsection{Single-mode photometers}
The focal plane contains four single-frequency channels; two at 90 GHz
and two at 150 GHz (Figure \ref{fig:feed}). The feeds are similar to 
those that will be used on the Planck Surveyor HFI (\cite{feed}).  The
dual-polarization feed structures are designed to provide low thermal
load on the  0.3~K stage, high efficiency and excellent control of
frequency and spatial response within a compact structure.

The entrance feeds are smooth-walled conical feeds designed to
illuminate the tertiary mirror with a -5~dB edge taper.  Using Gaussian
optics, the desired edge taper defines the beam waist at the phase
center of the horn, that in turn defines a relation between the horn
aperture diameter and the horn length.  The solution with minimum
length is selected, which gives diameters of 19.66~mm and 11.8~mm and
flare angles of 19.7$^{\circ}$ and 10.5$^{\circ}$  for the 90~GHz and
150~GHz structures respectively.

At the throat of the entrance feed, a $2\lambda$ length of waveguide
with a radius $r = 0.35\lambda$ defines the low frequency edge of the
passband and sets the throughput of the feed structure such that only
the lowest order Gaussian mode is propagated through the rest of the
system.   For the 90 and 150 GHz structures,  2.54~mm and 1.33~mm
diameter guides define cutoffs at 69~GHz and 132~GHz  respectively.

An f/4 re-expanding horn re-emits the light through band defining
filters and into an identical reconcentrating horn that feeds the light
into the detector cavity.  Identical low density polyethylene hyperbolic
lenses are placed at the aperture of each of the two face-to-face horns to
improve the coupling. The focal length of the lenses is set to maximally
match the lowest order Gaussian mode propagating from each horn through
the intervening filters in the thin lens approximation.

   The filter stack consists of two capacitive metal mesh filters 
(\cite{timusk}, \cite{filters}) and an alkali-halide/carbon layered 
polyethylene filter(\cite{yamada}).  The first capacitive
metal mesh filter defines the upper edge of the band 
while the second is tuned to a higher cutoff frequency and is used to 
block leaks in the first filter.  In the 150GHz feeds, filters with
cutoffs at 168 and 198 GHz are used, while in the 90 GHz feeds, 99 and
177 GHz filters are used.  The alkali-halide/carbon layered polyethylene 
filter blocks infrared and optical light ($\nu >$ 1650 GHz) and is tuned 
in thickness to maximize transmission at the center of the feed.

Each detector is mounted in an integrating cavity a distance $\lambda /
4$ from the exit aperture of the re-concentrating horn and at a distance
$\lambda / 4$ from a backshort in order to maximize the electric field
at the absorber (\cite{bolocavity}).  

We use foam eccosorb thermal loads at 300~K and at 77~K to measure
the optical efficiency of each of the feed structures.    The difference
in optical power is measured by comparing the I vs. V load curves of the
bolometer under the two different background conditions. We measure
optical efficiencies of 28\% for the 150 GHz  feed structure and 43\%
for the 90 GHz feed.   

 \subsubsection{Multi-color photometers}
The focal plane for the Antarctic LDB flight of \boom contains
4 three-color photometers (Figure \ref{fig:photometer}) similar to
those used in the FIRP instrument on IRTS (\cite{firp}),
MAX (\cite{max}), and  SuZIE (\cite{suzie_1.5}). These structures have
the advantage of simultaneous observation of the same region of sky in
three frequency bands. 

Back to back parabolic concentrators (\cite{winston}) mounted on the 2~K stage
feed the photometers.  The input horn is set at
f/3.4 to maximally couple to the re-imaging optics.  The horn has
an entrance aperture  diameter of 1~cm producing a 12$'$ beam on the sky,
corresponding to the diffraction  limit at 150 GHz.  The entrance aperture of the 
horn has a short flare in the shape of a $45^{\circ}$ section of a circle with a 
diameter of 1.3~cm to reduce sidelobes at low frequency. 

The throughput of the photometer is defined to be 0.05 cm$^2$ sr by the
1.45~mm diameter exit  aperture of the parabolic horn.   The
re-expanding parabolic horn has an aperture that matches a light pipe
located on the 0.3~K stage across a 0.5~mm gap.  At the entrance to the
light pipe, there is a metal mesh low pass filter with a cutoff of 480
GHz to improve the out-of-band rejection.

In the light pipe, two dichroic filters tipped 22.5$^{\circ}$ off-axis
direct light at frequencies greater than 270 GHz and 180 GHz through
inductive multilayer metal mesh bandpass filters that pass 410  GHz and
240 GHz respectively.  Radiation at frequency less than 180 GHz is
transmitted through the two dichroics and through a 150 GHz bandpass
filter.    Parabolic horns concentrate the light into integrating
bolometer cavities where it is detected. 

The three bolometers simultaneously observe the same part of the sky;
however the beam patterns of the three channels are quite different.  At 150
GHz, the entrance feed is single-moded, and by 410 GHz the system is
close to the geometrical optics limit.  The 150 GHz beam is
well-described by a Gaussian, and the 410~GHz the beam is close to a
top-hat.  At 240 GHz, several modes propagate in the horn, and the beam 
contains a mode which is annular in shape.

\subsubsection{Spectral Bandpass}

The spectral bandpass of the instrument is measured pre-flight using a
Fourier transform spectrometer (FTS).  Interferograms are measured for
each channel and Fourier transformed to measure the spectral response.
The final spectra are corrected for the 250 $\mu m$ thick polypropylene
beam splitter and the spectrum of the 77K thermal source.  The
bandpasses are plotted in figure \ref{fig:bandpass}.

Blocking each band with a high pass filter (thick grill filter) with a
cut-off frequency just above the band allows a search for leaks at high
frequency.  Since this test is sensitive to integrated leaks, a
much higher signal to noise measurement can be made than with the FTS. 
The signals produced by a chopped LN$_2$ source viewed with and without a
thick grill high pass filter in the beam were compared.   Table
\ref{tbl:blueleak} lists the results. 

\subsection{Calibration Lamp}

A 1~cm diameter hole located in the center of the tertiary mirror
contains a micromesh bolometer with copper leads for use as a
calibration lamp.   A germanium thermistor provides high impedance at
2~K. The electronics send a 1 second pulse of current through the
thermistor every 13 minutes, heating the bolometer and providing an
optical pulse to be used as a transfer calibrator.

\subsection{Attitude Control System}

The telescope is designed to smoothly scan in azimuth while at fixed
elevation, and has an attitude control system similar to those described
in \cite{bosca1}, \cite{bosca2}, and \cite{bosca3}.  The azimuth
pointing of the telescope is controlled by two torque motors (Inland
QT6205d); the first spins a large flywheel; the other torques
against the flight train of the balloon.  The two motors provide enough
torque to move the gondola in azimuth and to correct for random rotation 
of the balloon.

The elevation can be controlled by tipping the inner frame of the
telescope with a ball screw linear actuator (SKF CARN-32-300-1) driven
by a DC gearmotor (QT-1209-B).  Payload pendulations were reduced by an
oil-filled damper mounted near the flight train.  An overview of the
gondola frame is  shown  in Figure \ref{fig:gondola}.  The fully digital
control hardware consists of two redundant 386 CPU's.  A watchdog
circuit switches control of the pointing from one CPU in a few hundred 
milliseconds in case of reboot. 

Pointing information is provided by a differential GPS array, a two-axis 
sun sensor (\cite{sunsensor}), an encoder on the elevation axis, and three 
orthogonal axis laser
rate gyroscopes.   The azimuth gyroscope provides velocity feedback for
controlling the scan of the telescope.  The absolute pointing data from
the sun sensor is used to reset the drift in the gyroscopes and to
provide data for post-flight attitude reconstruction  

The telescope is powered by two solar power systems with lead-acid
battery backup.  The cryostat, receiver, and DAS are on a 300W 
system and the attitude control system and data storage are on a second
system with a 750W maximum.

\subsection{Thermal Design}

The outside of the gondola is paneled with 50 $\mu$m mylar bonded to 25
$\mu$m thick aluminum foil.  With the mylar side facing
outwards, the combination of  layers provides high reflectivity in the
optical band and high emissivity in the infrared band, thus protecting
the payload from excessive solar heating.

A thermal model is constructed for the payload that included heat
transfer by conduction and radiation alone.  Two cases were considered
for the input power: ``cold'' and ``hot''.  The cold case assumed that
the payload would drift over water and receive 1044 W m$^{-2}$ radiation
from the sun and a 9\% albedo for the surface of the earth for 177
W m$^{-2}$ of power.  The hot case assumed that the payload would be above
fresh snow and receive 1397 W m$^{-2}$ radiation from the sun and 213
W m$^{-2}$ from the earth with an albedo of 95\%.  Table \ref{tbl:thermal}
shows the predicted range of ambient temperatures.

\section{Observations}

\boom was launched from Williams Field at McMurdo Station
(Antarctica) on December 29, 1998 at 3:30 GMT on a 27 million cubic foot 
balloon with tied ducts and no radar tape.  The payload launch weight
was 1600 kg including 70 kg of glass ballast. The instrument reached
float altitude (38km) three hours later and began observations
immediately. The telescope remained within 1.5$^{\circ}$ of $\sim
78^{\circ}$ S latitude as it circumnavigated the continent. The flight
lasted 259 hours and was terminated 50 km south of the launch site.   

The majority of the flight was spent in CMB observation mode, that 
consisted of scanning the telescope in a smoothed triangle wave with a
60$^{\circ}$ peak- to-peak amplitude in azimuth at fixed elevation.
The center of the scan was set so that the region of sky away from the
galactic plane was mapped within the constraint of sun-avoidance.  As
the earth rotated, the scan center and  scan direction on the celestial
sphere changed.  At a single elevation, one day of scanning provided a
coverage of 22$^{\circ}$ in declination and gave scans tipped at $\pm
11^{\circ}$, providing cross-linking of the scans (Figure \ref{fig:scans}).  

Two  scan speeds were used in CMB mode; 1 degree per second (dps) and
2 dps. The elevation of the telescope was changed roughly
daily between angles of 40$^{\circ}$, 45$^{\circ}$, and 50$^{\circ}$.
Every 90 minutes, a 120$^{\circ}$ peak-to-peak scan centered on the
anti-sun direction was conducted for 5 minutes as a check for systematic
effects due to the sun.   Several HII regions in the galactic plane were
targeted during the flight as potential cross-calibrators and beam
mapping sources.  These  were RCW38, RCW57 (a double source composed of
NGC3603 and NGC3576), IRAS/08576, IRAS/1022, and the Carina Nebula.
Three known clusters (A3158, A3112, A3226) were targeted in a search for
the Sunyaev-Zel'dovich effect.  Three extragalactic point sources were 
observed serendipitously in the CMB map: the blazar 0537--441 (8544), the
BL Lac 0521--365 (8036), and the QSO 0438--443  (7044).   Table 
\ref{tbl:scanmodes} shows the total time spent in each scan mode. 

\section{Flight Performance}

\subsection{Detector System}

\subsubsection{Transfer Function}

A cosmic ray hit on a bolometer is well approximated by a delta
function power input, and can be used to measure the transfer
function of the bolometer and electronics.  The transfer function of the
experiment is  parameterized by the thermal time constant of the
bolometer and the properties of the anti-aliasing filters and AC-coupling
in the readout electronics.  The AC-coupling and anti-aliasing time
constants are measured on the ground and are expected to be the same in
flight since the electronics operating temperature in-flight is similar
to that on the ground.  However, the bolometer time constant is highly
sensitive to the background optical load. 

The combination of the theoretical transfer function of the bolometer
(a single pole low pass filter) and the measured electronics transfer
function is used to obtain the impulse response function.

A database of cosmic ray hits is built for each channel, simultaneously 
fitting an amplitude and phase shift to each hit as well as fitting the 
data to the impulse response function.    Figure \ref{fig:transfer}
shows the best fit template and the cosmic ray data for one of the 150
GHz channels.  The best fit bolometer time constants for each channel
are listed in Table \ref{tbl:performance}.  

The cosmic ray method is precise in the measure of the high
frequencies in the transfer function. The low frequency side of the
function is dominated by the 0.016~Hz AC-coupling filter. The best 
confirmation of the stability of the high pass time constants is made by 
the agreement of the calibration on the CMB dipole at the two scan 
speeds (Section \ref{calibration} below).

\subsubsection{Deglitching}

The \boom bolometer data are contaminated with transient events that
must be flagged and removed; these include cosmic ray hits, thermal
events in the 0.3~K stage, calibration lamp signals, and short periods
of electromagnetic interference (EMI).

Large thermal events in the 0.3~K stage and calibration signals appear
simultaneously in all channels and are easily flagged.  Smaller thermal
relaxation events were found with a pattern matching algorithm.  Cosmic
ray hits,  EMI spikes, and smaller thermal events occurring within an
individual bolometer are found using two algorithms.  First large spikes
are detected as deviations of greater than $3\sigma$ in a three point
difference function of the time ordered data, defined by $\delta_i = d_i
- 0.5 ( d_{i-2}+d_{i+2} )$, where d is the time-ordered sample and
$\sigma$ refers to the standard deviation of the data.  Smaller glitches
are found using an iterative binning scheme; the time stream data are
binned  into pixels on the sky and individual samples more than
$4\sigma$ from the average value of a pixel are flagged and not used in
the next iteration. After 4 iterations, a negligible number of new glitches are
found. 

The flagged data, identified by the methods given above, are replaced
by a constrained realization of the noise and not used in subsequent
analysis.  Approximately 5\% of the data in each channel are 
contaminated and flagged.  Drifts in the bolometer data after large cosmic
ray hits are removed by fitting a parabola to the data.  Drifts induced by
large thermal events are fit to an exponential and removed.  These data
are used in the subsequent analysis.

The average time between detected cosmic ray hits in each bolometer is
43 seconds. 

\subsubsection{Detector Noise}

The voltage noise is determined by taking the power spectrum of the
deglitched time domain data.  It is then deconvolved with the detector and
electronic transfer functions and converted to Noise Equivalent
Temperature (NET) by dividing by the DC detector responsivity.
Figure \ref{fig:noise} shows the NET as a function of frequency of a
150~GHz channel and Table \ref{tbl:performance} shows the average NET of
other detector channels.

During the 1 dps scans at 45$^{\circ}$ elevation, a signal on the 
sky with spherical multipole moment of $\ell=200$ is mapped to a
frequency of 0.38~Hz, well above the 1/f knee of the detector system. 

\subsection{Cryogenics}

The cryogenic system performed well, keeping the detectors well below
their required operating temperature of  0.3~K for the entire flight. 
Figure \ref{fig:temperatures} shows cryogenic temperatures recorded 
during the 10.5 day flight. The daily oscillation of the main helium 
bath temperature is due to daily
fluctuations of the external pressure. The altitude of the
payload varies with the elevation of the sun, which oscillates between
11$^{\circ}$ to 35$^{\circ}$ diurnally.  

We have searched for scan
synchronous temperature fluctuations in the $^3$He  
evaporator and in the $^4$He temperature, and we find 
upper limits of the order of 1 $\mu K_{rms}$ in the 1 dps 
scans, and drifts in the $^3$He evaporator temperature with an
amplitude of a few $\mu K$ during the 2 dps scans (Figure   
\ref{fig:thermal_yssn}).  These temperature fluctations are likely related to
the scan-synchronous noise observed during the 2 dps scans.

\subsection{Payload}

\subsubsection{Electronic and Thermal Performance}

The high cosmic ray flux above the Antarctic is a concern for digital
electronics.  Each of the two 486DX/4 based data storage
system computers experienced two watchdog-induced reboots during
the 259 hour flight.  The reboots were not concurrent,
so a continuous final data stream is reconstructed from the two systems.
The discrete CMOS based Data Acquisition System and the
80386 PC based Attitude Control System experienced no in-flight
lockups.

See Table \ref{tbl:thermal} for a comparison of predicted temperatures
with actual achieved temperatures.  The in-flight temperatures of the 
electronics were roughly 20$^{\circ}$C warmer than predicted,  perhaps due 
to white nylon blankets (not included in the model) placed over some of 
the electronics to avoid excessive cooling during ascent.

\subsection{Optics}

\subsubsection{In-flight Detector Background}

On day 8 of the flight, a partial I-V curve of each detector was
measured. The bias voltage was commanded to 4 discrete levels.   The
optical power  incident on the detector is determined by extrapolating
the in-flight I vs. V load curve to zero electrical power. Using
knowledge of the optical efficiency and the spectral bandpass, an
average equivalent Rayleigh-Jeans temperature of 10~K was found.  The
precision of the  background measurement is limited to $\sim 50\%$ by
uncertainty in the knowledge of the resistance  as a function of
temperature of the NTD germanium thermistors. 

\subsubsection{Beam Shape}
The intrinsic point spread function of each receiver is determined
by the combined effect of the telescope, the feed, the feed's
location in the focal plane, the spectral bandpass, the spectrum
of the source, and the distance to the source. In addition there
is a contribution to the effective beam on the sky associated with
error in the pointing reconstruction. To first order, this pointing
jitter manifests itself as an isotropic gaussian contribution to
the beam size.

In order to determine the physical beam, the optical system is
modeled with the ZEMAX ray tracing package (Focus Software, Inc.) as
well as a physical optics code written specifically for \boom (Figure   
\ref{fig:modelbeam}).  The 
results of the models are verified with near field beam maps made
with an ambient temperature eccosorb target.  The eccosorb target
has a Rayleigh-Jeans spectrum and allows a measurement of the beam
down the the $\sim$ -20 dB level (Figure \ref{fig:balloonball}). To 
measure the far field radiation pattern on the sky, single cuts through 
bright HII regions (primarily RCW38) are made during pointed observations 
of the Galactic plane.

Due to the extended nature of the Galactic sources it is necessary
to deconvolve the measured beam with the physical size of the
source in order to recover the far field radiation pattern.
Furthermore these sources do not exhibit the same spectrum as the
CMB anisotropy -- the Galactic sources observed by \boom 
exhibit a sharply rising spectrum consistent with 
a combination of free-free emission and thermal emission from dust 
(\cite{sources}). 
Nevertheless, scanned observations of bright HII
regions near the Galactic plane are used to estimate the beam full
width half maximum (FWHM) for each channel. These values are
presented in Table \ref{tbl:beams} and Figure \ref{fig:beam_summary}.

While the channels utilizing conical feed horns compare well with
the beam map data, a $\sim 10\% $ discrepancy is observed between
the calculated and measured FWHM of the photometer beams.  In all
cases, the measured FWHM are larger than the calculated beam size.
This discrepancy is attributed to the multi--moded nature of the
Winston concentrators utilized by these channels.  For these
channels the physical optics FWHM are scaled to fit the measurements of 
RCW38.

While these observations provide good statistics on the width of
the main lobe, the dynamic range of the measurements are not
sufficient to fully characterize the near sidelobe response and
deviation from gaussianity of the beam. To obtain a detailed model
of the near sidelobe response of the telescope, the physical
optics calculation of the beam shape is performed for each channel
using the measured position of the horns in the focal plane. The
two dimensional far-field radiation patterns from the physical
optics calculation scaled to fit the observed FWHM are then used to 
generate the window functions for each channel.

Observations of the position of the centroids of Galactic point sources 
provide a reliable way to quantify the contribution of the
pointing uncertainty to the effective beam size for CMB
observations.  The scatter in the centroids of these objects in maps 
produced with the final version of the pointing reconstruction (Section 
\ref{attitude}) indicate that there is a pointing uncertainty of 2.5$'$ 
($1 \sigma$).  

The final beam model used for CMB analysis consists of the ZEMAX
calculated beam (confirmed by the ground-based measurements) smeared
with the measured pointing jitter.

The far sidelobes in the 150~GHz channels were measured before flight
with a 150~GHz Gunn oscillator (Figure \ref{fig:sidelobes}).  Signal is
rejected at all angles greater than $\sim$5$^{\circ}$ at a level of
$<$-55 dB.  An upper limit to response from sources directly behind the 
telescope (eg the Sun) is set at $<$-82 dB.

\subsection{Attitude Reconstruction}
\label{attitude}

The attitude reconstruction is based on three sensors: the azimuth
sun sensor, a three axis rate gyro, and a differential GPS.  
The conversion from Azimuth-Elevation to Right Ascension-Declination 
additionally uses the time and position from a GPS receiver.

The azimuth sun sensor provides
a precise ($<1'$) and repeatable determination of the azimuth of
the gondola relative to the sun, but has a calibration that is dependent on 
sun elevation and sky background.  Additionally, 
the sensor is insensitive to rotations around the gondola-sun axis.

The rate gyros provide a three axis measurement of the motion of the
the gondola, but have a drift of 8$'$ hr$^{-1/2}$.

The differential GPS also provides a three axis measurement of attitude
of the gondola with good accuracy, but has an rms noise of 6$'$, and
suffered from a severe communications problem in flight with the pointing
computers in flight.

The azimuth sun sensor is calibrated by integrating the rate gyros over
a series of extended ($\pm$60$^{\circ}$) scans spaced throughout the flight at
different sun elevations, and then re-registering the offset by looking
at concurrent cleaned differential GPS data.

Once the azimuth sun sensor is calibrated, the telescope azimuth and 
elevation are determined by integrating the rate gyros with a 400s time 
constant.
The long term average of pitch and roll is set to zero, and the azimuth is
constrained on this time scale to fit the pitch and roll corrected azimuth
sun sensor.  Based on the gyro noise, we expect a 2.7$'$ ($1\sigma$) 
uncertainty.
Additionally, any long term pitch and roll offsets are filtered out by
this procedure.

We determine the offset of each detector and the rms noise on the sky relative to the
gondola attitude from sub-maps made from in-flight
observations of galactic point sources.  This analysis  
revealed an elevation offset correlated with the gondola temperature 
which is regressed out.  The final scatter in the residual of the 
galactic plane point source centroids is 2.5$'$ ($1\sigma$).

\section{Calibration}
\label{calibration}

\subsection{CMB Dipole}

The 90, 150, and 240 GHz bolometer signals are calibrated by their
response to the CMB dipole anisotropy. The CMB dipole is an ideal
calibration source for a CMB mapping receiver.  Its amplitude has been
measured to high precision (0.7\%) by COBE/DMR (\cite{cobecal}). Its
spectrum is identical to that of the degree-scale anisotropy and it
entirely fills the \boom beams. This minimizes the final calibration
error by  eliminating uncertainty in the measured spectral frequency 
response of each channel and in the beam pattern of the telescope. 

The direction of the CMB dipole is roughly orthogonal to the \boom scan
direction and it is observed continuously during the CMB scan mode.  
The dipole appears as a 3 mK peak-to-peak signal in the
timestream, much larger than the detector noise.

The dipole signal appears at f=0.008Hz and f=0.016Hz during the 
1 dps and 2 dps CMB scan modes respectively.  At such low
frequencies the detector signals are sensitive to 1/f noise,
uncertainties in the measurement of the transfer function, and scan
synchronous noise.  To mitigate the worst of these while still
retaining the large scale information in the dipole, the time-ordered
data are high-pass filtered with a filter described by 
$F=0.5\left(1-\cos\left(\pi\nu\over f_0\right)\right)$ for $0<\nu<f_0$
and 1 elsewhere, where $f_0$ is set to 0.01~Hz.

To calibrate, we artificially sample the CMB dipole signal
 (\cite{Line1996}), including a correction for the Earth's velocity around the sun
 (\cite{Stump1980}), according to the \boom scanning, and filter this
fake time stream in the same way as the flight data. 
The 1 dps flight data are then fit to the model.

The galactic plane lies at one edge of the map, and bright galactic
dust in the plane can create a spurious dipole signal. To check for
Galactic contamination, various Galactic cuts are made, and the fits
are expanded to include a model of dust emission (\cite{cmbdust}),
filtered as above. The results are insensitive to these changes to
within the quoted uncertainties, indicating that Galactic emission does
not contaminate the calibration.

During the CMB scan mode, an additional scan synchronous signal from
an unknown source appeared at the fundamental scan frequency.  In the
150~GHz and 240~GHz channels during the 2 dps mode, this
appeared as a signal larger than the signal from the dipole (Figure
\ref{fig:TOD}).  The scan synchronous signal is well correlated with
the 410~GHz channels, and a 410~GHz map is used as a template to model
the noise. The fits were again extended to include the possibility of
a component correlated with the 410~GHz data, and again the results
do not change to within the quoted uncertainties. This indicates
that the calibration is insensitive to a wide range of
systematics such as atmospheric contamination, as these would be
traced by the 410~GHz data.

Removal of the scan-synchronous noise limits the precision of the dipole
calibration.  A conservative measure of how well the scan-synchronous
noise is removed is 10\%,  the level of agreement between the
1 dps and 2 dps calibrations. 

\subsection{Galactic Point Source Cross Calibration}

Point source observations were made in January 2000 of the three
extragalactic sources and images were made of the sources NGC3576 and
RCW38 at 90 and 150GHz with the SEST telescope at LaSilla Observatory in
Chile. The size of the SEST beam is 57$''$ FWHM at 90 GHz and 35$''$ FWHM
at 150 GHz. The SEST data is convolved with the \boom beam to
check the calibration of the the 150 and 90 GHz \boom channels. Details
and source fluxes are given in \cite{sources}.  

Maps of NGC3576 and RCW38 centered on the core of the source (4$' \times$4$'$ in
extent) were made with 17.5$''$ spacing and extended maps (10$'
\times$10$'$ in extent for NGC3576, 6$' \times$6$'$ in extent for RCW38)
were made with 35$''$ spacing.  We use a combination of 
these maps for RCW38 to calculate fluxes integrated to 1 and 2 $\sigma$ 
points of the \boom beam, and the center map only to compute the 
integrated flux of NGC3576.  These 1 and 2 $\sigma$ fluxes are dependent 
on beam shape, but independent of flux background.

The SEST predicted fluxes are compared with the fluxes found inside the 
1 and 2 $\sigma$ contours of the \boom data.  These provide a cross-check
on the primary \boom calibration, which is derived from the CMB dipole.
We find that the 150 GHz galactic source fluxes are in agreement 
with the dipole calibration to roughly 6\%, while the 90 GHz dipole
calibration is approximately 8\% higher than expected given the SEST
data.  

\subsection{Responsivity calibration of the 410 GHz channels}
Due to the higher background, the 410 GHz channels are less
sensitive than the lower frequency channels. They also feature greater 
scan-synchronous noise. The CMB dipole, which is detected at the
scan frequency, is significantly contaminated by spurious signals,
and cannot be used as a calibrator. For these channels we 
use a different calibration method, based on the degree-scale
anisotropy of the CMB, which is present in the 410 GHz maps as
well as in the lower frequency maps. This is detected at a
frequency much higher than the scan frequency, and a sharp
high-pass filter can be used to remove scan-synchronous noise 
and drifts in the scan direction, while retaining the CMB signal.
The pixel averaging present in the mapping process efficiently
dilutes the effect of scan synchronous noise in the direction
orthogonal to the scan. We use the maximum likelihood sky
maps obtained with the iterative algorithm of \cite{simon}.
Samples of these maps have been published in \cite{pdb}.

We correlate each high-pass filtered 410 GHz map (which is a
combination of noise, interstellar dust emission and CMB
anisotropy) with the sum of a high-pass filtered 150 GHz map and a
high-pass filtered interstellar dust template. The former is
dominated by CMB anisotropy: the interstellar dust signal at 150
GHz is less than 1$\%$ of the detected mean square fluctuations
(\cite{boomdust}). The latter monitors the ISD component present in
the 410 GHz map, and is obtained from the IRAS/DIRBE maps
(\cite{dustmap}) of interstellar dust emission. It is filtered
in the same way as the \boom signals (\cite{boomdust}). The CMB
anisotropy is thus the main signal in common between the 150 GHz
map and the 410 GHz map, expressed as:
\begin{equation}
\label{eqn:dust}
V_{410} = A \Delta T_{150} + B I_{dust} + C 
\end{equation}
where $V_{410}$ is the map of the 410 GHz signal to be calibrated,
expressed in Volts; $\Delta T_{150}$ is the map of the 150 GHz
signal, expressed in $\mu K_{CMB}$; $I_{dust}$ is the dust
template. The coefficient $A$ can be estimated from a linear
regression and is the gain calibration factor (responsivity) for
the 410 GHz channel, expressed in $V \mu K_{CMB}^{-1}$. If the
IRAS/DIRBE map is properly extrapolated and its brightness is
expressed in $\mu K_{CMB}$, the best fit value for $B$ is also an
independent estimate of the responsivity of the 410 GHz channel.

We divide the observed sky into many $10^o \times 10^o$
patches, and for each patch where we have more than 100 pixels in
common between the 3 maps we perform a best fit using Equation 
\ref{eqn:dust}, leaving $A$, $B$ and $C$ as free parameters. We  
average the
best fit coefficients found for each sky patch, and estimate the
error in the final coefficients as the sum in quadrature of the
standard deviation of the best fits plus the calibration error in
the 150 GHz map.

We do not include any estimate of the systematic 
effects in the extrapolation of dust in the uncertainty in $B$. There
is a galactic latitude dependance of the results for $B$, probably
because low latitudes could be contaminated by cold dust clouds that are not
accurately modelled by the IRAS/DIRBE map.

The typical total calibration error for all 410 GHz channels
is $\sim 30\%$, dominated by the presence of systematic effects
(the statistical uncertainty on individual fits is less than $10\%$).

We test the stability of the calibration over the course of the flight by dividing the data into two halves.  The calibration in each of the two halves of the flight is stable to well within the total calibration error.

\subsection{Calibration Stability}

Drifts in the responsivity are measured by response to the calibration
lamp.  In every channel,  the drift in responsivity is well-modeled by a 
linear drift that is removed from the final calibrated time stream.
The average drift in responsivity over the entire 10.5 day flight is 3.6\%. 

\subsection{Sensitivity}

The NEP of each detector is determined by dividing the voltage noise by
the electrical responsivity obtained from an I vs. V load curve measured
on the ground.  The responsivity of the detector is scaled to flight
load conditions via the bolometer DC voltage.

The sensitivity of the \boom receiver to CMB fluctuations is determined
by dividing the measured voltage noise by the responsivity to the
dipole.  The average NEP and NET$_{CMB}$ for each frequency  is listed
in Table \ref{tbl:performance}. 

The optical efficiency of each channel is determined by comparing the
predicted responsivity to the CMB dipole to that measured in flight.
The predicted responsivity of a channel is given by $\frac{dV}{dT} =
\frac{dV}{dP} A \Omega \int \frac{dB_\nu (2.7K,\nu)}{dT} t(\nu) d\nu$ 
where $\frac{dV}{dP}$ is the electrical responsivity measured from the
bolometer I-V curve, $A\Omega$ is the photometer throughput ,
$\frac{dB_\nu}{dT}(2.7K,\nu)$ is the spectrum of the CMB dipole,  and
$t(\nu)$ is the passband of the photometer channel.  The average
optical efficiency for each frequency is listed in Table
\ref{tbl:performance}. 

\section{Conclusion}

The \boom payload performed to specification in its first long
duration flight above Antarctica.  The cryogenic system held the
detectors below 0.285~K for the duration of the 10.5 day flight. The
attitude control system and the data acquisition performed
flawlessly. Ambient temperatures on the payload stayed within operating range.

The receiver performed well, achieving an average instantaneous
sensitivity of 140 $\mu$K$\sqrt{\rm s}$ in the 90 and 150~GHz single
mode channels.  The micromesh design for the bolometer absorbers limited
the cosmic ray contamination to 5\% of the data.  The quasi total power
radiometer provided stability over a wide enough range of signal
frequencies to allow mapping of roughly 2000 square degrees of sky, a
calibration to 10\% from the CMB dipole, and precision measurement of
CMB anisotropies on degree scales.  

The cosmological results from the Antarctic flight of this instrument are 
reported in \cite{pdb}, \cite{b98parm}, \cite{pdb02}, and \cite{cbn2}.

\smallskip

The \boom project is supported by the CIAR and NSERC in Canada; by 
PNRA, Universit\'{a} ``La Sapienza'' and ASI in Italy; by PPARC in the UK; 
and by NASA, NSF, OPP, and NERSC in the US.  The authors would like to 
thank Kathy Deniston for logistical support, and NASA's National 
Scientific Balloon Facility (NSBF) and the US Antarctic Program for 
excellent field and flight support.

\clearpage


{\par\centering \resizebox*{0.7
\columnwidth}{!}{\rotatebox{270}{\includegraphics{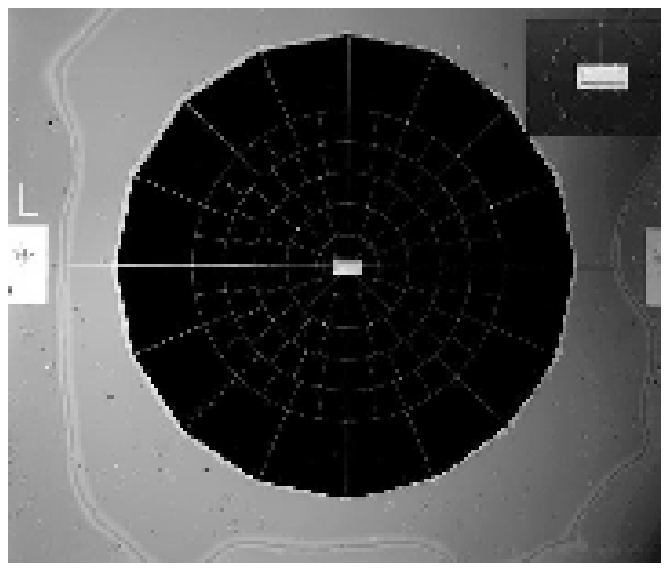}}} \par}
\vspace{1cm}
\figcaption[Lweb.eps]{
A ``spider-web'' bolometer used in \boom.  The micromesh absorber is 4mm 
in diameter.  The NTD-Ge thermistor is visible in the center of the
micromesh absorber, and is shown in the inset.
\label{fig:bolopic}}

{\par\centering \resizebox*{0.7 
\columnwidth}{!}{\includegraphics{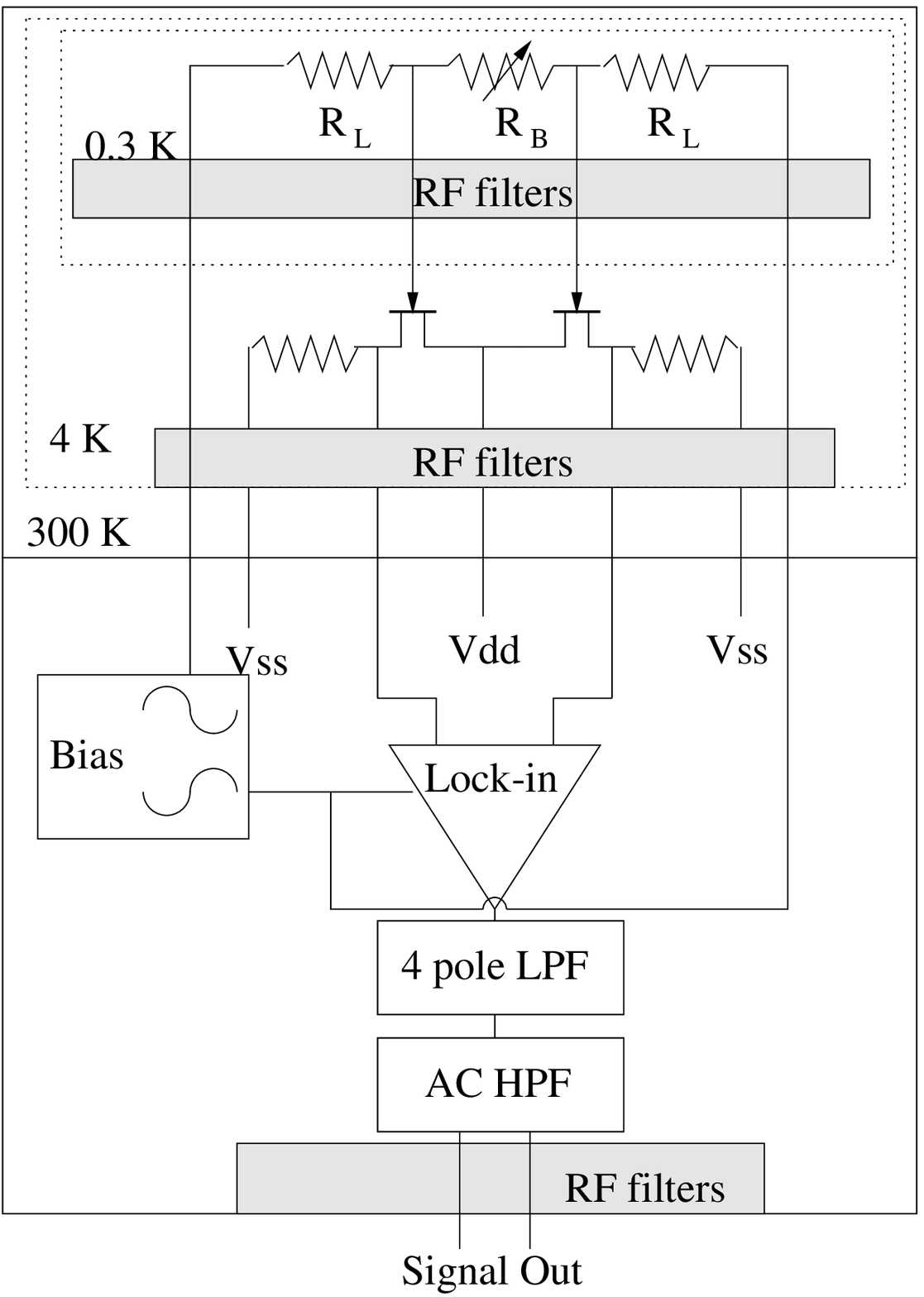}} \par}
\figcaption[readout.eps]{
A block diagram of the bolometer readout electronics.  A symmetric 331~Hz sine wave 
biases each bolometer across two 10 $M\Omega$ load resistors R$_L$.  A 
matched dual JFET pair lowers the output impedance, and sends the detector 
signals to the warm electronics.  There, the two signals are differenced, 
amplified, and synchronously demodulated.  The lockin output is filtered 
and sent via a differential output to the data acquisition system.
 \label{fig:readout}}

\smallskip
{\par\centering \resizebox*{1 \columnwidth}{!}{\includegraphics{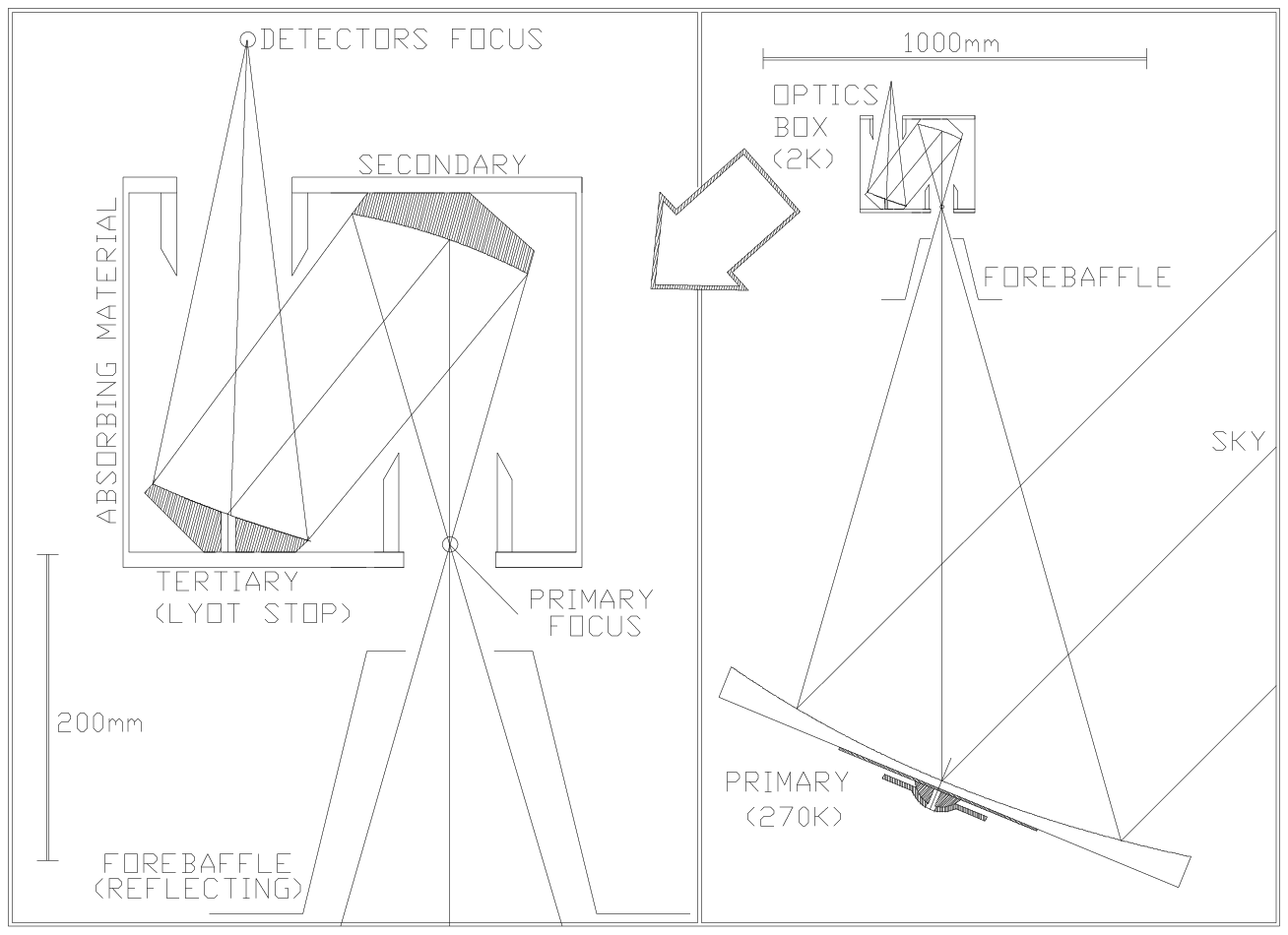}} \par}
\figcaption[figureoptics.eps]{
The \boom optics.
\label{fig:optics}}
\smallskip

\smallskip
{\par\centering \resizebox*{1 \columnwidth}{!}{\includegraphics{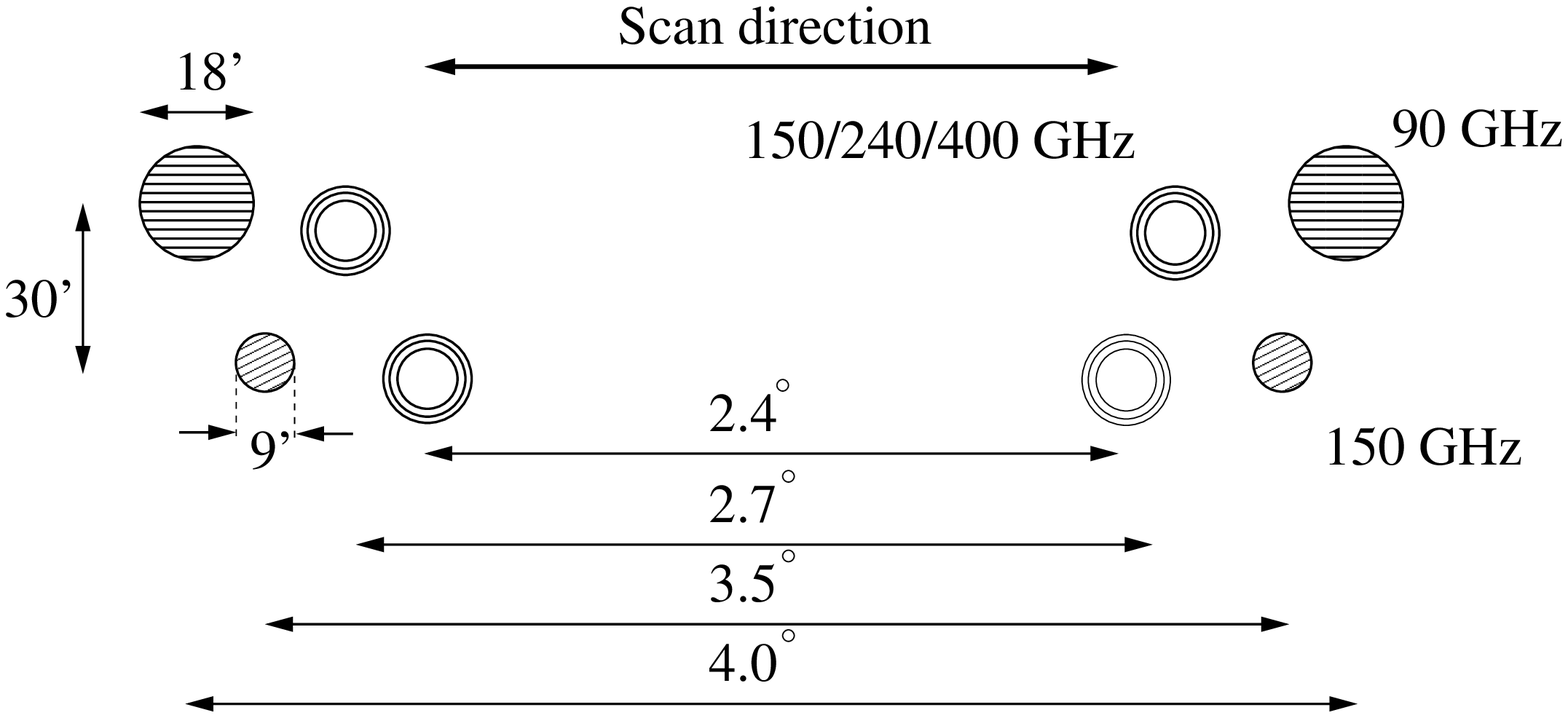}} \par}
\figcaption[focalmap.eps]{
The \boom focal plane projected onto the sky.  The multi-color photometer 
beams are roughly 9$'$, 14$'$, and 12$'$ for the 150,240, and 410 GHz 
channels respectively.
 \label{fig:focalmap}}
\smallskip

\smallskip
{\par\centering \resizebox*{1 \columnwidth}{!}{\includegraphics{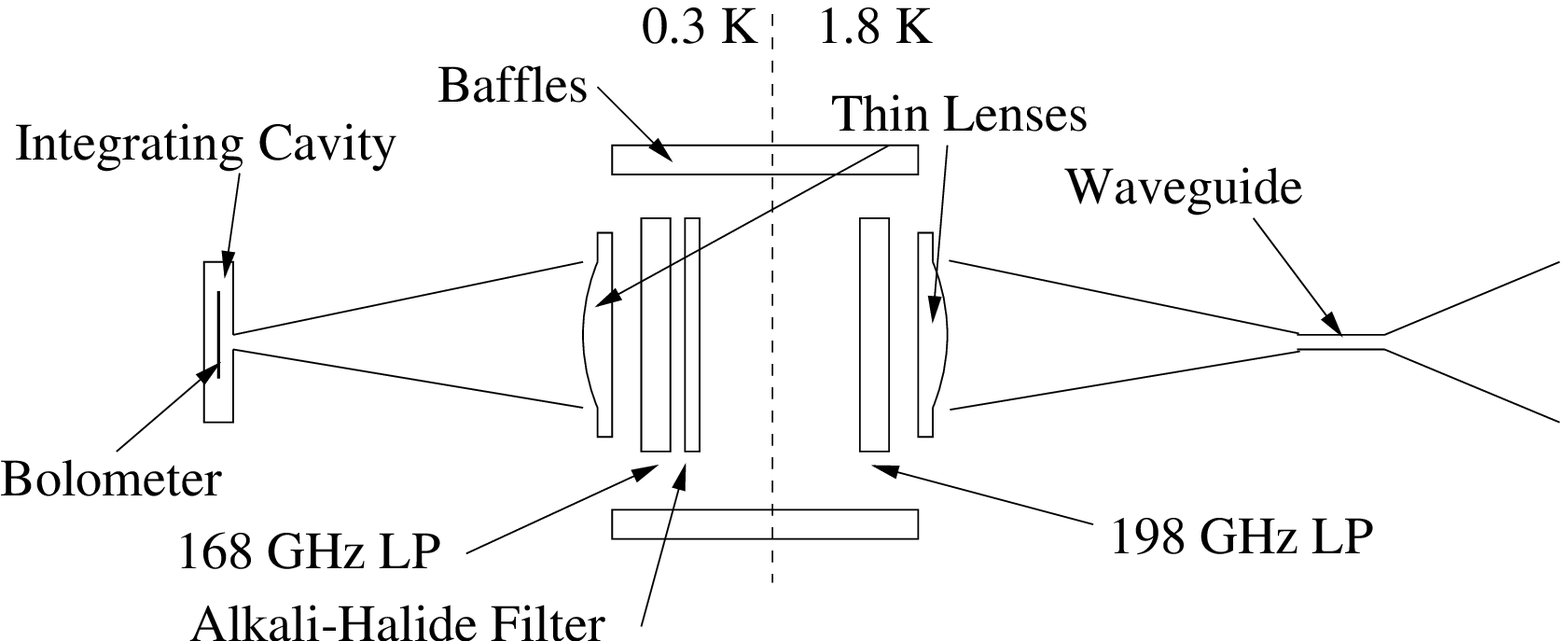}} \par}
\figcaption[feed/feed2.eps]{
A schematic of a \boom 150 GHz single frequency feed.  The 90 GHz feed
structures are similar in design. 
\label{fig:feed}}
\smallskip

\medskip
{\par\centering \resizebox*{0.8 \columnwidth}{!}{\includegraphics{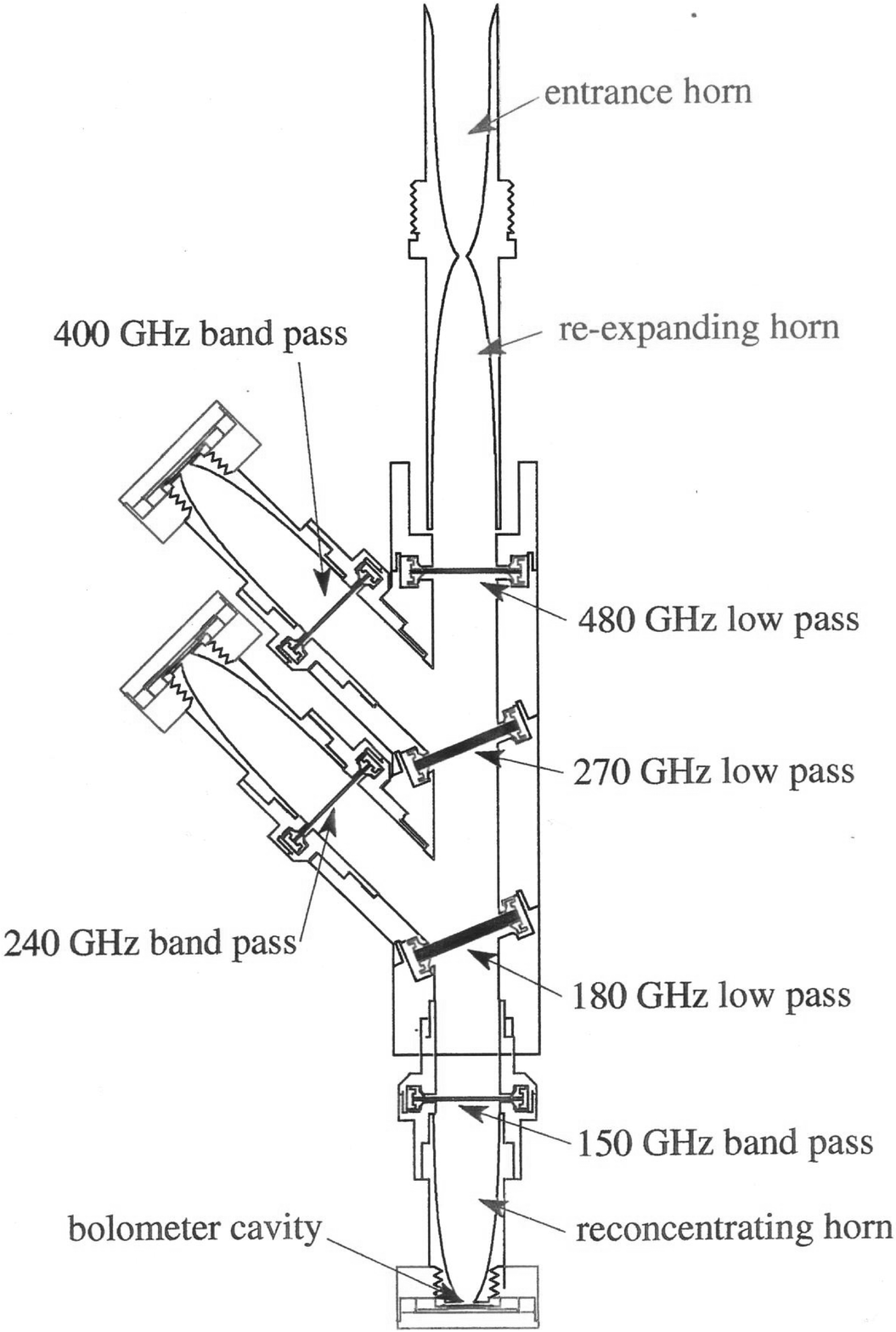}} \par}
\figcaption[photometer.eps]{\small
A \boom multi-color photometer.
\label{fig:photometer}}

\medskip
{\par\centering \resizebox*{1 \columnwidth}{!}{\includegraphics{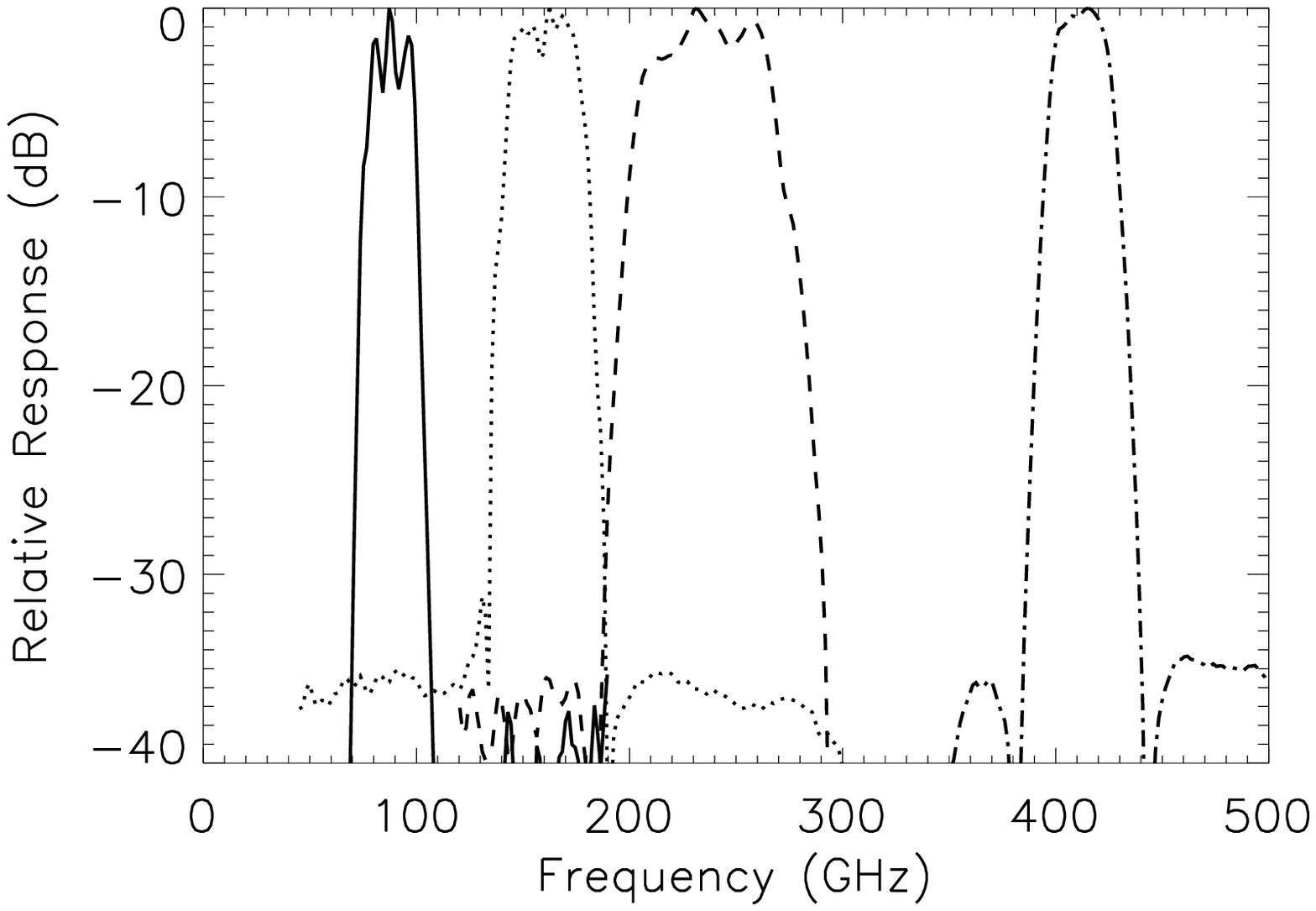}} \par}
\figcaption[LDBbands.ps]{
Relative spectral repsonse of four channels to a flat-spectrum source.
The peak response in each band has been normalized to unity.  The signal 
at -35 dB in each band is rectified noise.
\label{fig:bandpass}}
\medskip

\medskip
{\par\centering \resizebox*{1 \columnwidth}{!}{\rotatebox{90}{\includegraphics{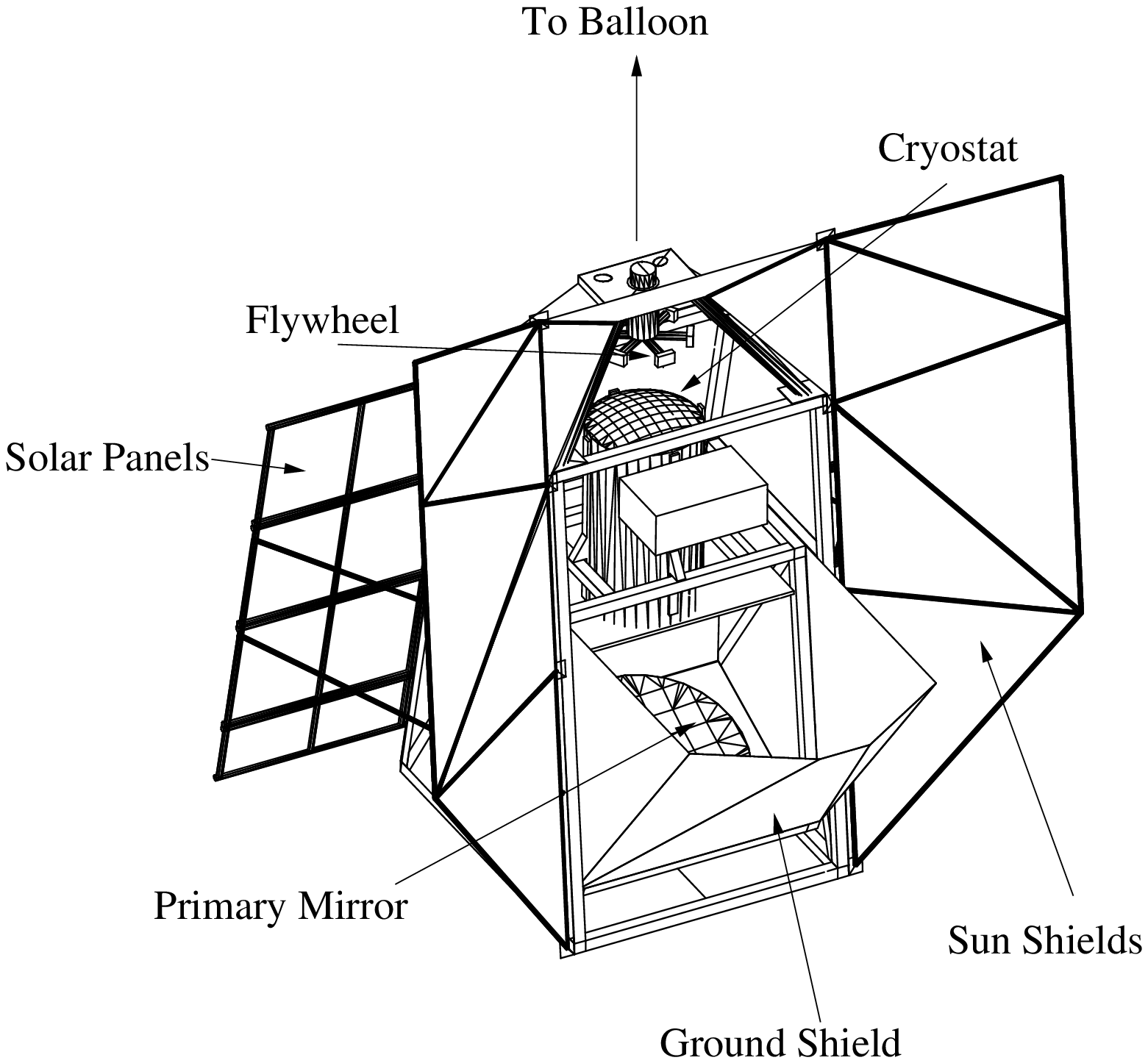}}}
\figcaption[gondola.eps]{
The \boom gondola.  The height of the payload is roughly 4 m from the
bottom to the heighest point on the sun shields.  The mass of the
payload at launch, including ballast, is 1600 kg.  The center sun shield 
is oriented vertically to avoid the reflection of ground radiation to the
telescope. 
\label{fig:gondola}}
\medskip

\medskip
{\par\centering \resizebox*{1 \columnwidth}{!}{\includegraphics{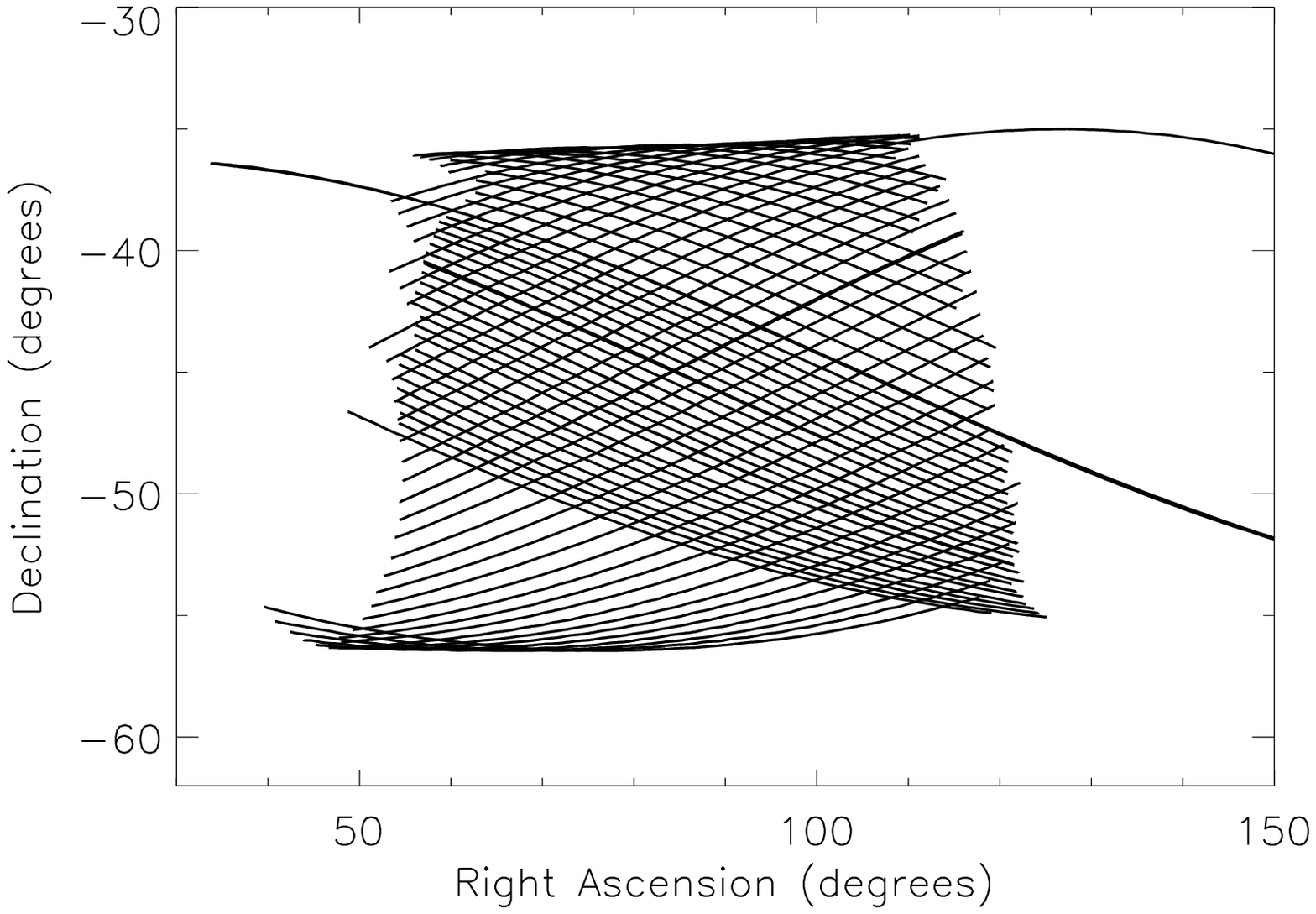}}
\figcaption[scans.ps]{
The scan crosslinking of the \boom flight.  Every 20th scan from a 22.5 hour 
section on days 4 and 5 of the flight is shown.  The telescope is at a fixed 
elevation of 45$^{\circ}$. During the first five hours of this section of the 
data, the telescope scans at 2 dps, yielding the series of 
closely spaced scans at the center of the plot.  During the remainder of 
the section, the telescope scans at 1 dps.  This section of 
the flight is chosen for clarity; observations immediately before and 
after this section are made at other telescope elevations.
\label{fig:scans}}
\medskip

\medskip
{\par\centering \resizebox*{1 \columnwidth}{!}{\includegraphics{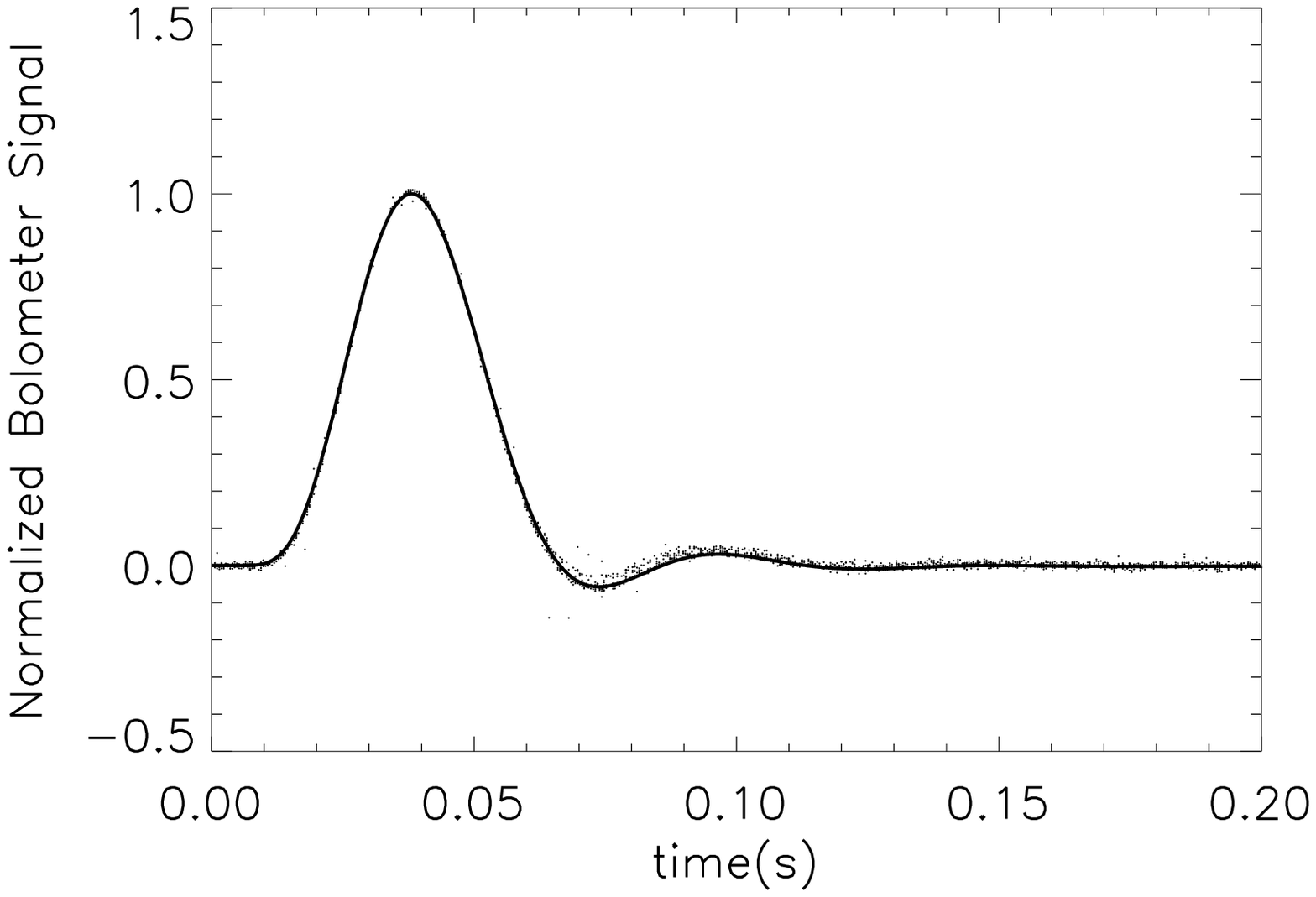}} \par}
\figcaption[transferfn.ps]{
The best fit impulse response function to cosmic ray hits for one of the 
150 GHz bolometers.  The dots in the figure show all of the time stream
data of cosmic ray hits, phase shifted to the best fit model.
\label{fig:transfer}
}
\medskip

\medskip
{\par\centering \resizebox*{1 \columnwidth}{!}{\includegraphics{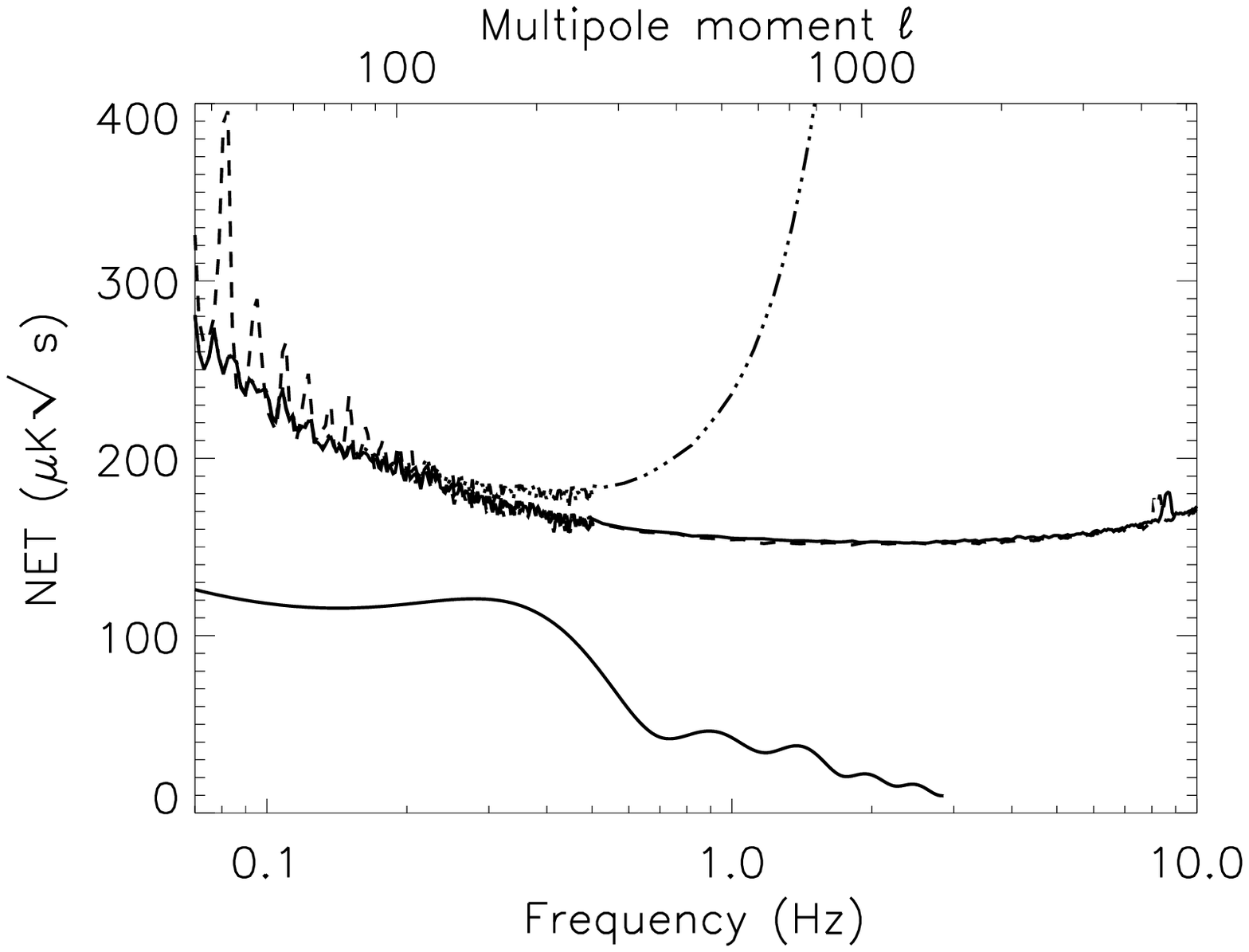}} \par}
\figcaption[LDBnoise.ps]{
Two noise spectra of a 150 GHz \boom detector during 1 dps mode
(solid line) and 2 dps mode (dashed line).  The lines at
$f < 0.1$ Hz are harmonics of the scan frequency.  The rise in NET at high
frequency is due to the bolometer's time constant $\tau = 10.8$ms.  The
top x-axis shows the corresponding spherical harmonic multipole for the
1 dps mode.    The dot-dashed line shows the 1 dps noise spectrum 
convolved with the window function of a 10$'$ beam.  
The solid line shows the expected signal due to CMB anisotropy.
\label{fig:noise}}
\medskip

\medskip
{\par\centering \resizebox*{0.9  
\columnwidth}{!}{\rotatebox{0}{\includegraphics{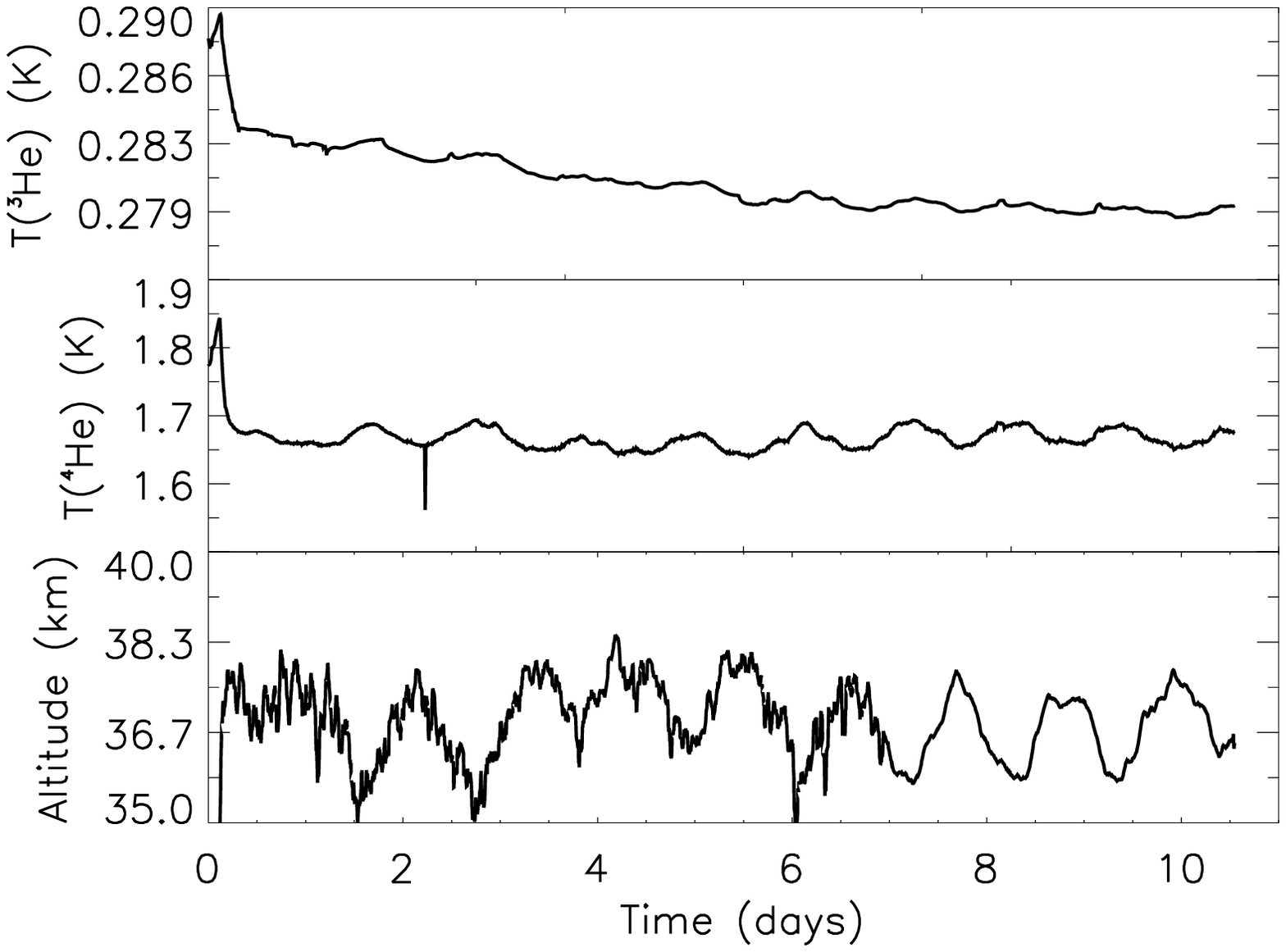}}}
\par} 
\figcaption[temperatures.ps]{
The temperature of the $^4$He cold stage and of the $^3$He evaporator
during the flight.  The  lower panel shows the altitude of the payload
(measured using the onboard GPS) that anticorrelates well with both
cryogenic temperatures. 
\label{fig:temperatures}}
\medskip

\medskip
{\par\centering \resizebox*{1.0 
\columnwidth}{!}{\rotatebox{0}{\includegraphics{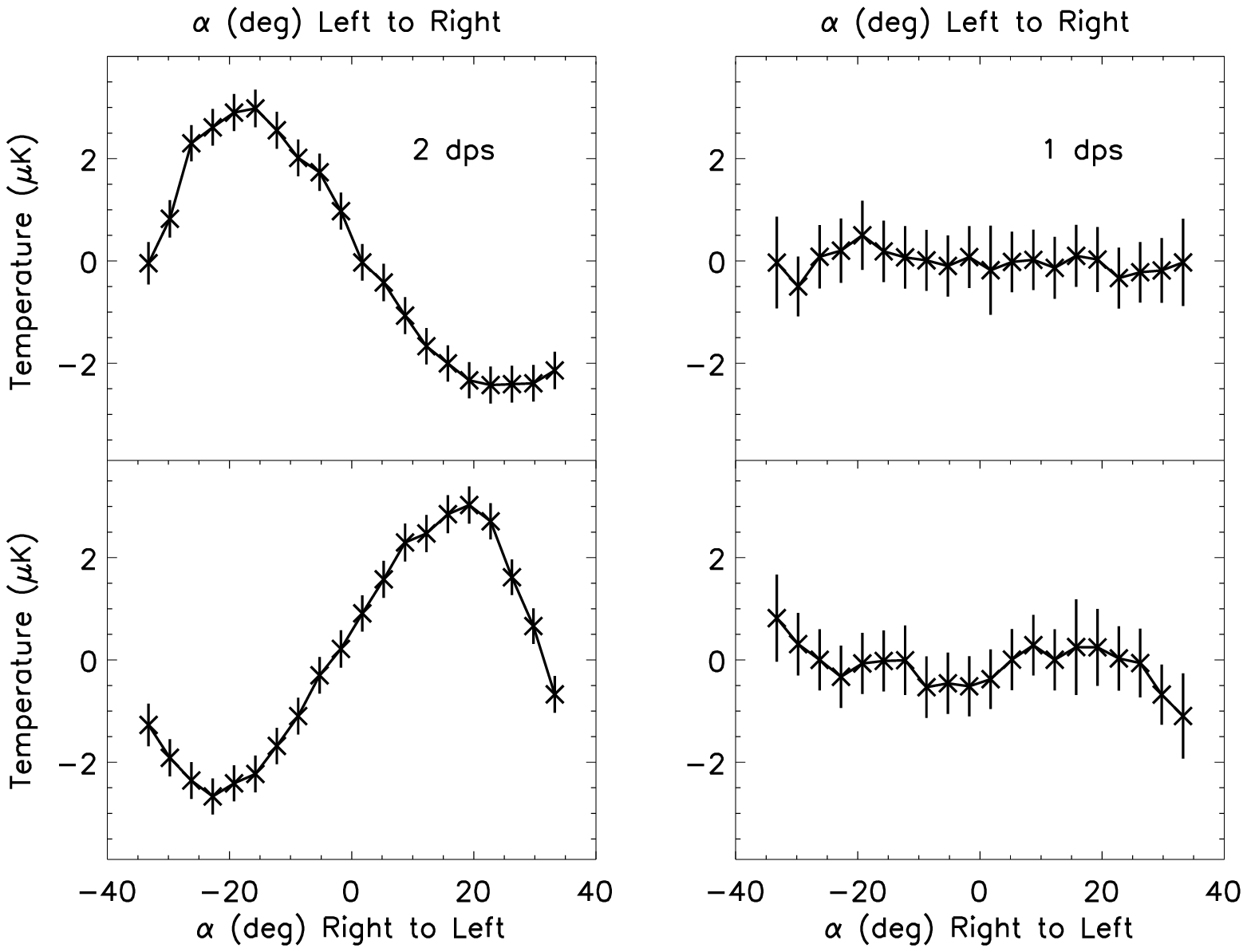}}}
\par}
\figcaption[thermal_yssn.ps]{
Fluctuations of the temperature of the $^4$He cold stage, binned in angle 
relative to the center of the telescope scan.  The top row contains data 
from the left-to-right direction of the scans, and the bottom row contains 
data from the right-to-left direction of the scans. The right column 
shows 1 dps data and the left column contains 2 dps data. 
Each plot consists of 24 hours of data.
\label{fig:thermal_yssn}} \medskip

\medskip
{\par\centering \resizebox*{1 \columnwidth}{!}{\includegraphics{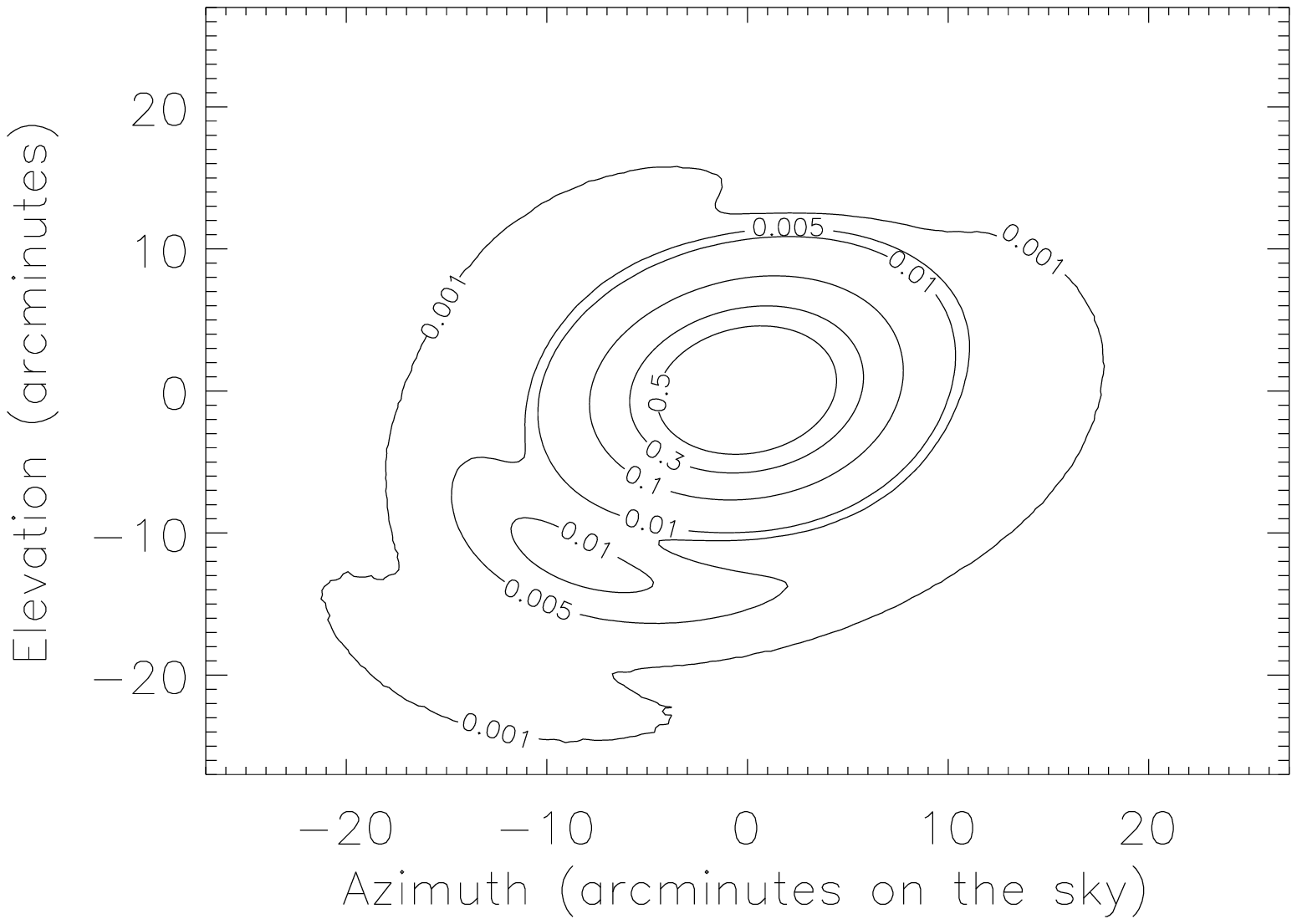}} \par}
\figcaption[150Azmax.ps]{
ZEMAX model for one of the 150 GHz beams, computed with a band-averaged
point spread function for a gaussian illumination.
\label{fig:modelbeam}}
\medskip

\medskip
{\par\centering \resizebox*{1 \columnwidth}{!}{\includegraphics{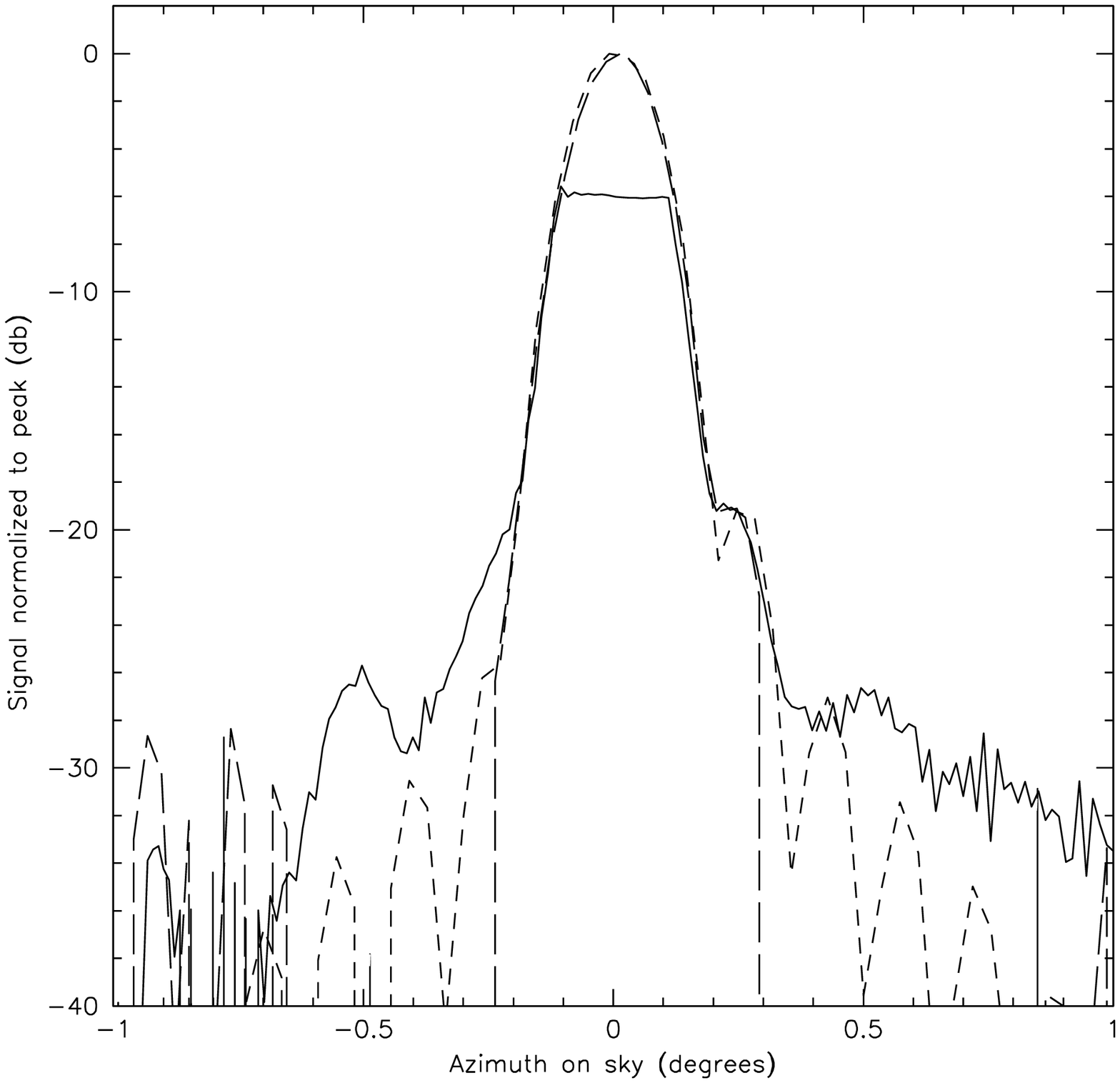}} \par}
\figcaption[balloonball.ps]{
150 GHz data from one scan in azimuth across the tethered
sources.  The solid line shows the  signal from the large source, and is 
saturated in the center of the beam.  The long-dashed line shows the signal
from the small source. The short-dashed line shows the ZEMAX model. 
\label{fig:balloonball}}
\medskip

\medskip
{\par\centering \resizebox*{1 \columnwidth}{!}{\includegraphics{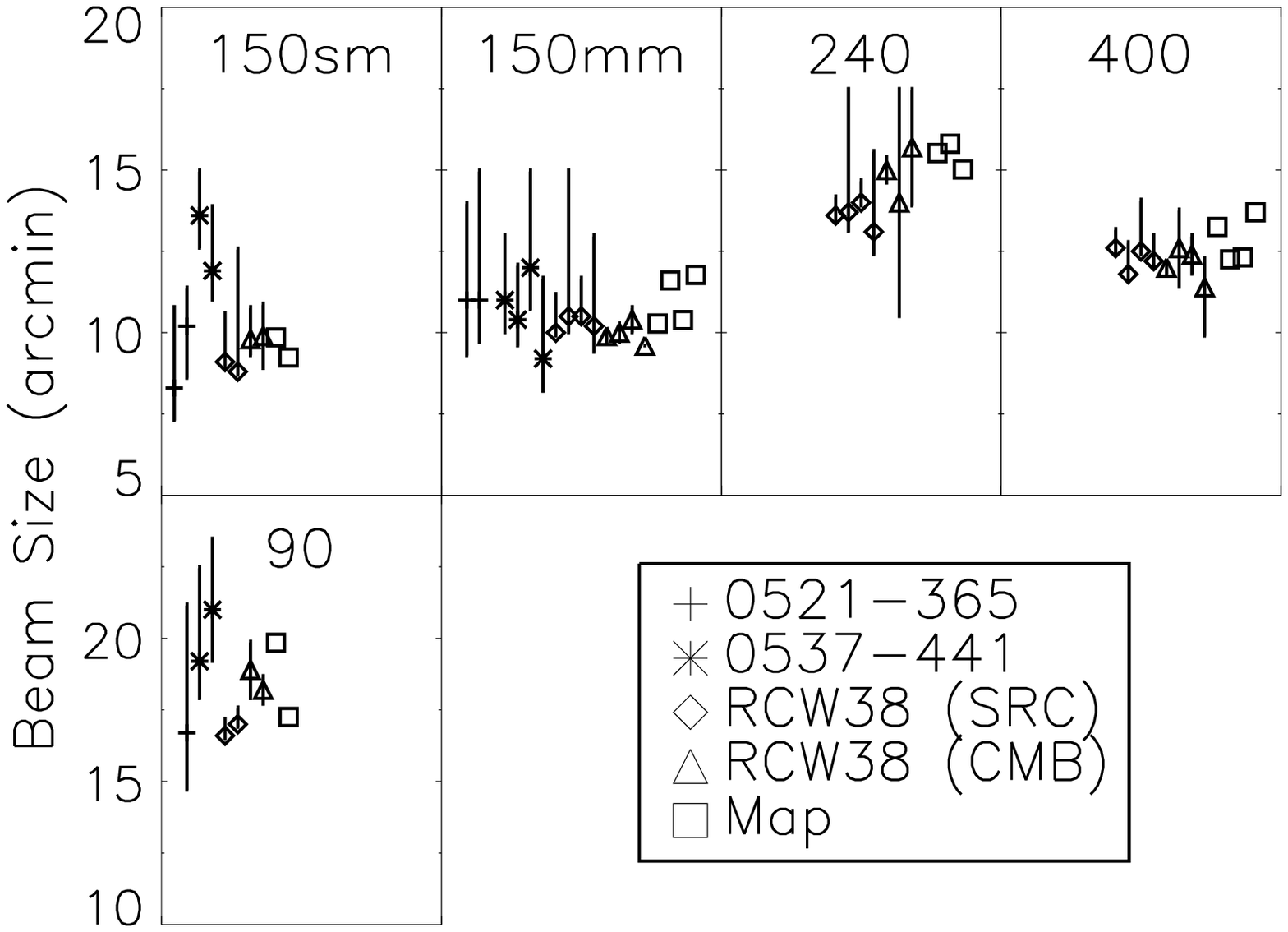}} \par}
\figcaption[beam_summary_p.ps]{
Summary of the beam FWHMs for all the channels using the three in-flight
methods.  ``0521--365'' and ``0537--441'' are the beam sizes derived from the radial
profile of the two brightest extragalactic point sources. ``RCW38 (SRC)'' is
the beam size derived from the targeted observation of RCW38, and ``RCW38
(CMB)'' is from the serendipitous observation of RCW38.  ``Map'' is the
beam size derived from the 2D binned source maps.  These measurements
are used only to confirm the ZEMAX+jitter model; poor signal to noise
and uncertain source profiles prevent the use of the in-flight
measurements for a beam map.  The 150 GHz channels are divided into 
single-mode (150sm) and multi-mode (150mm).
\label{fig:beam_summary}}
\medskip

\medskip
{\par\centering \resizebox*{1 \columnwidth}{!}{\includegraphics{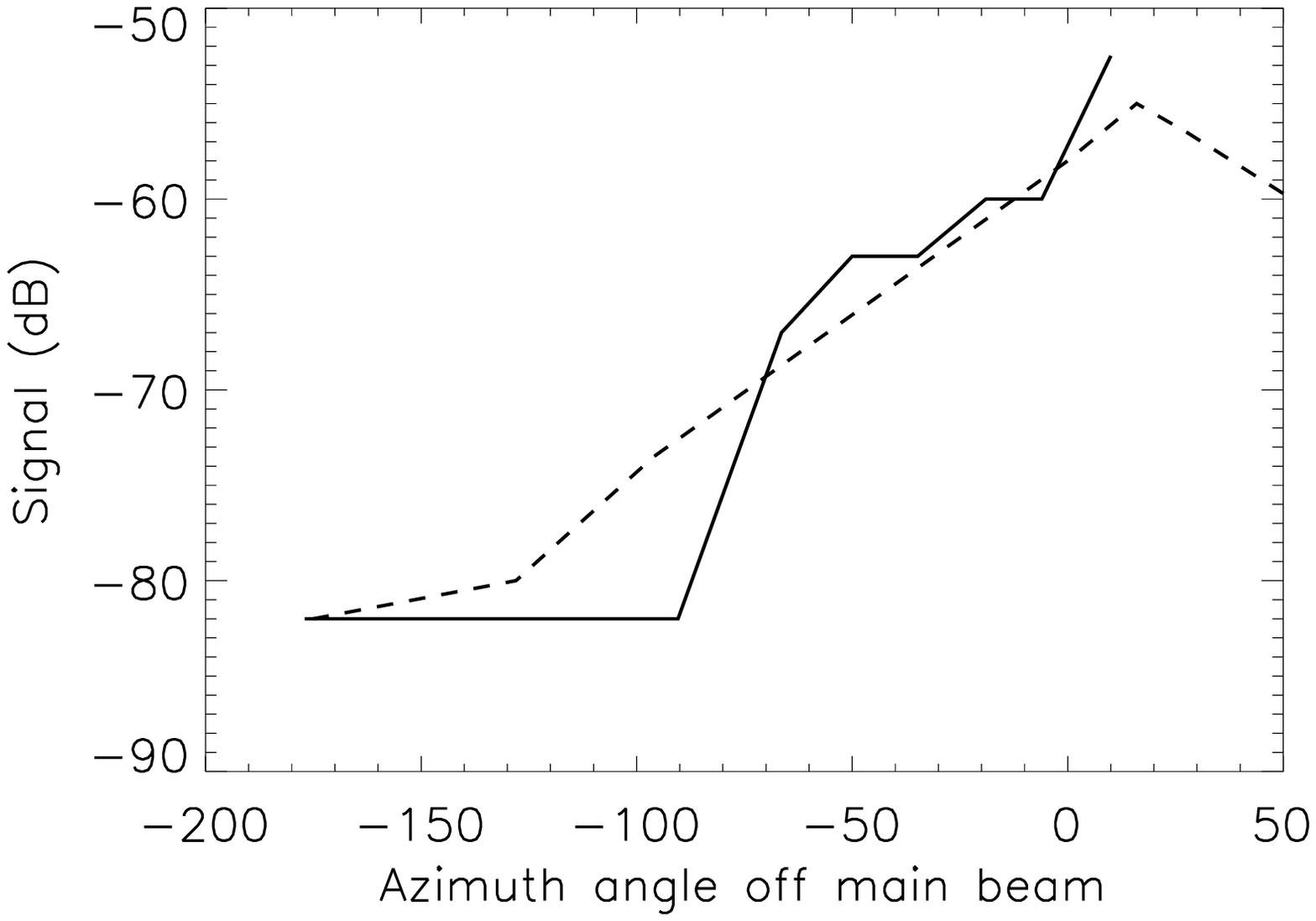}} \par}
\figcaption[sidelobes.ps]{
A measurement of the far sidelobes of the \boom telescope with the 150
GHz Gunn oscillator normalized to the peak on-axis response.  The solid
line shows the sidelobes with the source at 33$^{\circ}$ elevation and
the dashed line shows the sidelobes  with the source at 60$^{\circ}$
elevation.  The noise floor is at -82 dB.
\label{fig:sidelobes}}
\medskip

\medskip
{\par\centering \resizebox*{1 \columnwidth}{!}{\rotatebox{90}{\includegraphics{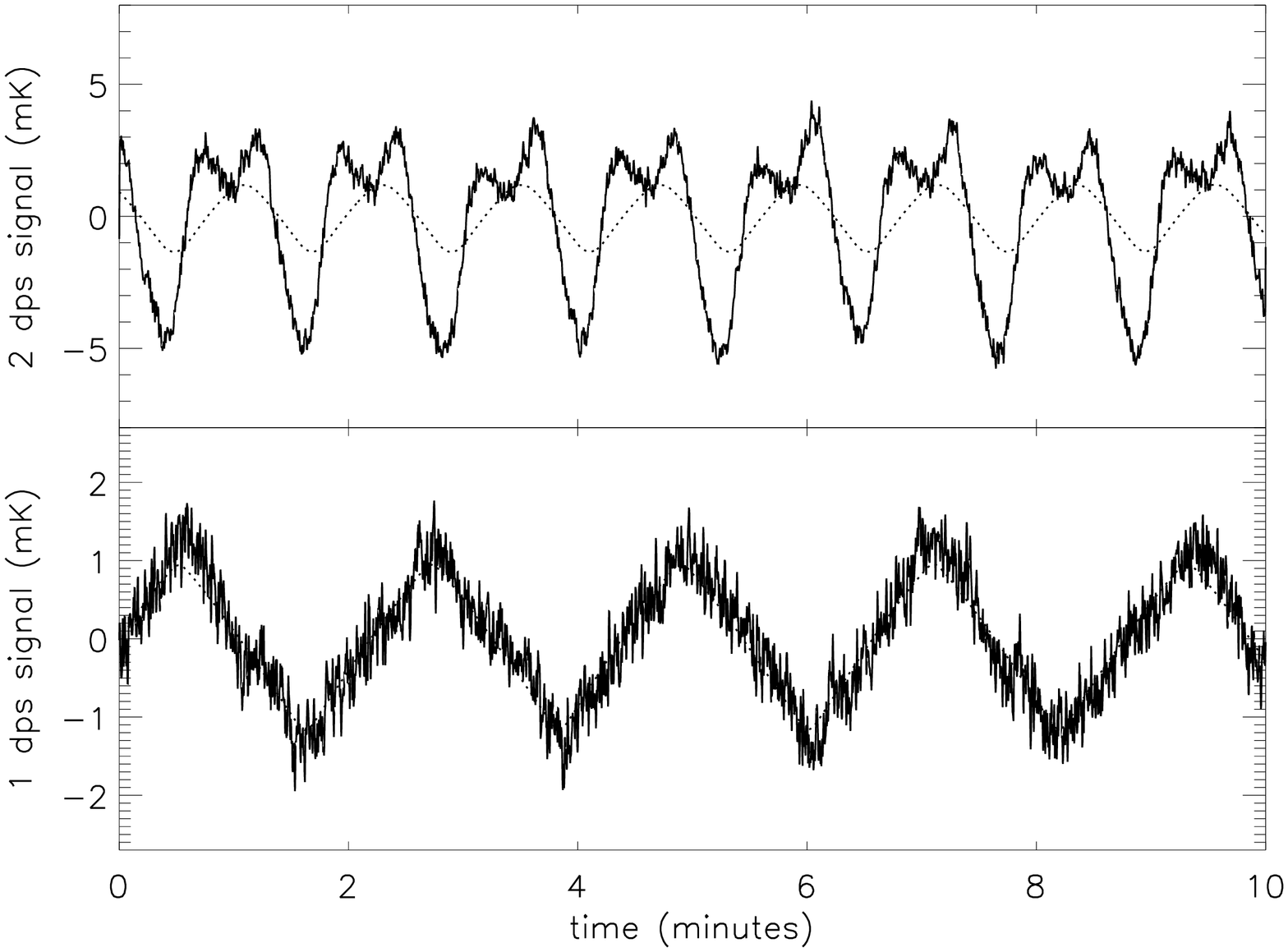}}} \par}
\figcaption[boomTOD.ps]{
A sample section of deglitched, calibrated time ordered data for a 150
GHz detector for each of the CMB scan modes.  The bandwidth has been
limited to 2 Hz to show the low frequency signal.  The best fit CMB dipole
is overlaid in each panel as a solid grey line. In 2 dps 
mode, the additional scan synchronous signal dominates the dipole, whereas 
the 1 dps signal is dominated by the CMB dipole.
\label{fig:TOD}
}
\medskip


\clearpage

\begin{deluxetable}{cccc}
\tabletypesize{\scriptsize}\tablecaption{High Frequency Spectral Leaks\label{tbl:blueleak}}
\tablewidth{0pt}\tablehead{\colhead{$\nu_0$ (GHz)}&\colhead{Thick Grill
Cutoff (GHz)}&\colhead{Out-of-band Power (R-J)}&\colhead{Out-of-band
Power (dust)}} \startdata
90  & 150 &  1.7\% &  23\%\\
150 sm& 230 & 0.5\% & 3.1\%\\
150 mm& 230 & 0.9\% & 5.7\%\\
240 & 300 &  0.8\% & 2.6\%\\
410 & 540 &  0.5\% & 0.8\%\\
\enddata
\tablecomments{The ratio of out-of-band to in-band power for a source with 
a Rayleigh-Jeans spectrum (column 3) and a dust spectrum (column 4).  A 
result much less than 100\% in column 4 indicates that the dust 
sensitivity of that channel is given simply by the in-band dust emission.
The results shown are the average over channels at each frequency.  
Because of different filtering schemes, the 150 GHz channels are divided 
into single-mode (sm) and multi-mode (mm) here.  The above band leak is
presumed to produce a flat response from the cutoff of the highpass
filter to the cutoff of the alkali-halide filter at 1650~GHz.  A single
component dust model with emissivity spectral index $\alpha$ = 1.7 and dust
temperature  $T_{dust}$ = 20~K is used (\cite{cmbdust}).}

\end{deluxetable}

\begin{deluxetable}{ccc}
\tabletypesize{\scriptsize}\tablecaption{Ambient Thermal Performance 
\label{tbl:thermal}}\tablewidth{0pt}\tablehead{\colhead{Item} &
\colhead{Predicted Temperature ($^{\circ}$C)} & \colhead{Measured
Temperature ($^{\circ}$C)}} \startdata
Attitude Control System  & -8$^{\circ}$ to 12$^{\circ}$&  15$^{\circ}$  to 30$^{\circ}$  \\
Data Storage System& 17$^{\circ}$ to 27$^{\circ}$ & 33$^{\circ}$  to 42$^{\circ}$ \\
Data Acquisition System& -7$^{\circ}$ to 6$^{\circ}$ &  18$^{\circ}$  to 29$^{\circ}$ \\
Cryostat & -29$^{\circ}$ to 27$^{\circ}$ &  -5$^{\circ}$  to 13$^{\circ}$  \\
Bolometer Readout Electronics& -31$^{\circ}$ to 2$^{\circ}$ & 21$^{\circ}$  to 27$^{\circ}$ \\
Solar Array & 55$^{\circ}$ to 92$^{\circ}$ & 57$^{\circ}$ to 68$^{\circ}$ \\
Ground Shield & no prediction & -37$^{\circ}$ to -17$^{\circ}$ \\
Primary Mirror & no prediction & -12$^{\circ}$ to 1$^{\circ}$ \\
Gondola Frame & no prediction &  15$^{\circ}$ to 28$^{\circ}$\\
\enddata
\tablecomments{A comparison of the payload thermal model with the temperature 
achieved in flight.  The two predicted values are for the ``cold'' and
``hot'' cases.  The two measured values are the minimum and maximum
temperatures reached during the daily cycle.  The discrepancy in 
predicted and in-flight temperatures for the electronics is likely due to 
white nylon blankets placed over the electronics to avoid excessive 
cooling during ascent.} 
\end{deluxetable}

\begin{deluxetable}{cc}
\tabletypesize{\scriptsize}\tablecaption{Flight Scan Modes
\label{tbl:scanmodes}}\tablewidth{0pt}\tablehead{\colhead{Target} &
\colhead{Time (hr)} } \startdata
1 dps CMB &  105.8\\
2 dps CMB &  82.0 \\
WIDESCAN & 10.7\\
\tableline
IRAS/08576 & 9.2\\
IRAS/1022 & 2.7\\
RCW38 & 7.4\\
RCW57 & 1.4\\
$\eta$ Car & 3.3\\
\tableline
Cen A & 2.2\\
\tableline
A3158 & 9.9\\
A3112 & 3.3\\
A3226 & 6.4\\
\tableline
Diagnostics & 8.9\\
\enddata

\tablecomments{Time in hours spent in various scan modes during the \boom
flight. The first two modes are CMB scan mode.  ``WIDESCAN'' refers to
the $\pm$60$^{\circ}$ scans.  IRAS/08576, IRAS/1022, RCW38, RCW57, and
$\eta$ Car refer to targeted observations of galactic sources. Cen A
refers to an attempted targeted observation of an extragalactic source.
A3158, A3112, and A3226 are observations of known Abell clusters.
Diagnostics consists of time spent trimming the bias level and spinning
the payload.}

\end{deluxetable}

\begin{deluxetable}{cccc}
\tabletypesize{\scriptsize} \tablecaption{Instrument
Characteristics. \label{tbl:beams}} \tablewidth{0pt} \tablehead{
\colhead{Channel} & \colhead{Band (GHz)}   & \colhead{$NET_{CMB}$
($\mu K \sqrt{s}$)}   & \colhead{Beam FWHM (')} } \startdata
B150A  & 148.0 - 171.4 & $130$  & $9.2\pm 0.5$\\
B150B  & 145.8 - 168.6 & Variable & $9.2\pm 0.5$\\
B150A1 & 145.5 - 167.3 & $231$  & $9.7\pm 0.5$\\
B150A2 & 144.0 - 167.2 & $158$  & $9.4\pm 0.5$\\
B150B1 & 144.2 - 165.9 & $196$  & $9.9\pm 0.5$\\
B150B2 & 143.7 - 164.3 & $184$  & $9.5\pm 0.5$\\
\tableline
90 (2 Chs)   &  79  -  95  & $140$  & $18\pm 1  $\\
240 (3 chs) & 228  - 266 & $200$  & $14.1\pm 1$\\
410 (4 chs) & 400  - 419 & $\sim 2700$ & $12.1\pm 1$\\
\enddata

\tablecomments{Summary of relevant instrument characteristics.
Only results from the 150GHz channels are presented in this paper.
B150B is not used due to non-stationary detector noise. The
bandwidth limits are computed to include 68\% of the total
detected power for a flat spectrum source. The NET is computed at
1Hz.}
\end{deluxetable}

\begin{deluxetable}{ccccccc}
\tabletypesize{\scriptsize} \tablecaption{In-flight bolometer
performance\label{tbl:performance}} \tablewidth{0pt} \tablehead{
\colhead{$\nu_0$} & \colhead{$\tau$} & \colhead{$\eta_{opt}$} &
\colhead{G}  & \colhead{R} & \colhead{NEP (1 Hz)} &
\colhead{NET$_{CMB}$}\\ (GHz) & (ms) &  & (pW K$^{-1}$) &(M$\Omega$) &
($10^{-17}$  W/$\sqrt{\rm Hz}$) 
&($\mu$K$\sqrt{\rm s}$)}
\startdata
90  & 22 & 0.30 & 82 & 5.5 & 3.2 & 140\\
150sm & 12.1 & 0.16 & 85 & 5.9 &  4.2 & 140\\
150mm & 15.7 & 0.10 & 88 & 5.5 & 4.0 & 190 \\
240 & 8.9 & 0.07 & 190 & 5.7 & 5.7 & 210\\
410 & 5.7 & 0.07 & 445 & 5.4 & 12.1 & 2700\\
\enddata

\tablecomments{
In-flight bolometer performance.  The 150 GHz channels are divided into
single mode (150sm) and multimode(150mm).  The optical efficiency of the 
channels decreased significantly from the measured efficiency of each
feed structure due to truncation by the Lyot stop. The NEP is that
measured in flight, and includes contributions from detector noise,
amplifier noise, and photon shot noise.
}
\end{deluxetable}

\end{document}